%% file: main.tex
\documentclass[sigconf,authorversion,screen,nonacm]{acmart}
\acmConference[ICSE 2022]{The 44th International Conference on Software Engineering}{May 21–29, 2022}{Pittsburgh, PA, USA}

\usepackage[utf8]{inputenc}
\usepackage{url}
\usepackage{tabularx}
\usepackage{amsmath,amsfonts}
\usepackage[noend]{algpseudocode}
\usepackage{amsthm}
\usepackage{booktabs}
\usepackage{multirow}
\usepackage{graphicx}
\usepackage{textcomp}
\usepackage{tikz}
\usetikzlibrary{topaths,calc,shapes.misc,shapes}
\input{tikz_sets}
\usepackage{xcolor}

\usepackage{enumitem}
 \usepackage{listings}
\usepackage[linesnumbered,ruled,lined,noend]{algorithm2e}
\SetArgSty{textnormal}
\usepackage{graphicx}
\usepackage{hyperref}
\usepackage{subfig}

\hypersetup{
  colorlinks=true,
  citecolor=Violet,
  linkcolor=Red,
  urlcolor=Blue}

\usepackage{multirow}
\usepackage{listings}
\usepackage{color, soul}
\definecolor{gray}{RGB}{215,215,215}
\sethlcolor{gray}
\usepackage{textcomp}
\usepackage{framed}
\usepackage{hhline}
\usepackage[breakable]{tcolorbox}
\usepackage{float}
\usepackage{array}
\usepackage[keeplastbox]{flushend}
\usepackage{makecell}
\usepackage{balance}
\usepackage{pifont}
\usepackage{xspace}
\usepackage{tcolorbox}
\usepackage{microtype}
 
\input{macro}

\newcommand{\ToolName}{BeDivFuzz}

\title{\textsc{\ToolName}: Integrating Behavioral Diversity into Generator-based Fuzzing}

\copyrightyear{2022}
\acmYear{2022}
\setcopyright{acmcopyright}\acmConference[ICSE '22]{44th International Conference on Software Engineering}{May 21--29, 2022}{Pittsburgh, PA, USA}
\acmBooktitle{44th International Conference on Software Engineering (ICSE '22), May 21--29, 2022, Pittsburgh, PA, USA}
\acmPrice{15.00}
\acmDOI{10.1145/3510003.3510182}
\acmISBN{978-1-4503-9221-1/22/05}

\begin{document}
\begin{abstract}
A popular metric to evaluate the performance of fuzzers is branch coverage.
However, we argue that focusing solely on covering many different branches (i.e., the \textsl{richness}) is not sufficient
since the majority of the covered branches may have been exercised only once,
which does not inspire a high confidence in the reliability of the covered code.
Instead, the distribution of the executed branches (i.e., the \textsl{evenness}) should also be considered.
That is, \textsl{behavioral diversity} is only given if the generated inputs
not only trigger many different branches, but also trigger them evenly often with diverse inputs.
We introduce \textsc{\ToolName}, a feedback-driven fuzzing technique for generator-based fuzzers.
\textsc{\ToolName} distinguishes between \textsl{structure-preserving} and \textsl{structure-changing} mutations
in the space of syntactically valid inputs,
and biases its mutation strategy towards validity and behavioral diversity
based on the received program feedback.
We have evaluated \textsc{\ToolName} on Ant, Maven, Rhino, Closure, Nashorn, and Tomcat.
The results show that \textsc{\ToolName} achieves better behavioral diversity than the state of the art,
measured by established biodiversity metrics, namely the Hill numbers, from the field of ecology.
\end{abstract}

\author{Hoang Lam Nguyen}
\affiliation{%
  \institution{Humboldt-Universität zu Berlin}
  \country{Germany}}
\email{nguyehoa@informatik.hu-berlin.de}

\author{Lars Grunske}
\affiliation{%
  \institution{Humboldt-Universität zu Berlin}
  \country{Germany}}
\email{grunske@informatik.hu-berlin.de}


\maketitle

\section{Introduction}\label{sec:introduction}

Traditionally, fuzzing tools (e.g., \cite{afl, Bohme2016AFLFast, libFuzzer, honggfuzz, peachFuzzer, Wang2017Skyfire, RawatJKCGB17VUzzer})
have been used to evaluate the software under test (SUT)
with respect to security and robustness properties.
Typically, vulnerabilities are found by feeding the SUT malformed
inputs, potentially resulting in unexpected program behavior,
which can be identified using e.g., memory and safety oracles \cite{NagarakatteZMZ09,SerebryanyBPV12,StepanovS15MemorySanitizer}.
Since most of the vulnerabilities emerge due to incorrect handling of unexpected inputs,
security-oriented fuzzers usually target the input parsing and processing stages of the SUT.
Recently, there is a trend~\cite{AschermannSAH20,reddy2020rlcheck,Padhye2019Zest,ToffolaSP17,YeTTHFSBW021,ZongLWDL020} to use fuzzers
to test the actual core functionality of the SUT,
rather than the early input processing stages only.
The challenge of testing the core functionality of a program
that expects complex structured inputs is mainly due to the following problems:
\begin{enumerate}
	\item The input must be \textsl{syntactically} valid (i.e., conform to the expected structure/type) in order to be parseable by the SUT.
	\item The input must satisfy any additional \textsl{semantic} validity constraints (e.g., assertions or \texttt{repOk} methods~\cite{LiskovG86}) to actually reach the core program functionality.
	\item The generated inputs must exhibit some sort of \textsl{diversity} in order to trigger diverse behavior.
\end{enumerate}

The techniques targeting this problem typically
rely on an input specification (e.g., a grammar)
that describes the expected input structure to
produce inputs of the expected format.
Generator-based fuzzers like Zest~\cite{Padhye2019Zest} follow an imperative approach,
where the tester implements a \textsl{generator} program that is able to
produce syntactically valid inputs.
Zest uses code coverage and validity feedback to
search for inputs that exercise many different branches in
the semantic analysis stages of the SUT.
That is, the goal of Zest is to \textsl{cover} as much of the semantic program behavior as possible.
A more recent technique, RLCheck~\cite{reddy2020rlcheck}, utilizes reinforcement learning to
automatically guide the generator towards high input diversity, i.e.,
inputs that exercise different traces.
However, while RLCheck is able to
produce a large number of diverse inputs that trigger \textsl{specific behaviors},
it fails to cover many different behaviors.

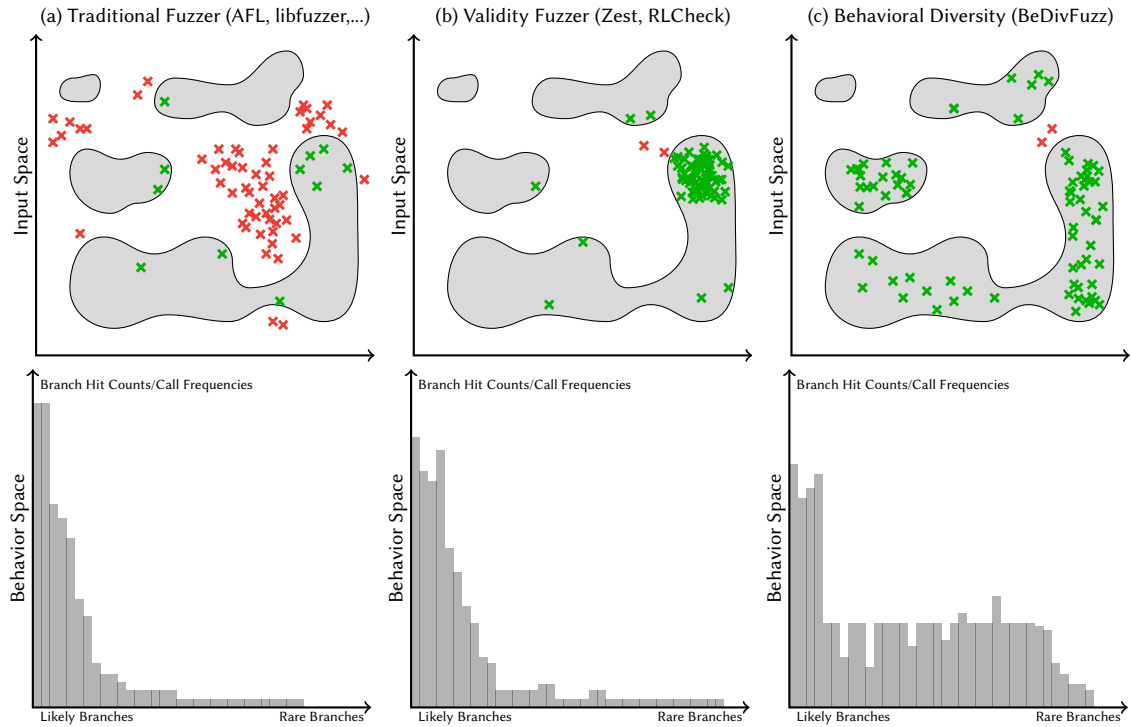
\begin{figure*}[t]
  \centering
  \input{figure_1}
  \caption{
    Illustrative simplified motivation for \textsc{\ToolName}.
    In the input space, the white and grey areas represent invalid and valid input areas for the SUT, respectively.
    The green crosses represent valid test inputs, while the red crosses represent invalid inputs. In line with \cite{soremekun2020inputs}, we use branch hit counts and call frequencies as an illustrative metric to represent the behavior space.
  }
  \label{fig:motivation}
\end{figure*}

We are interested in a technique that is able to not only \textsl{cover}
many different behaviors,
but also test these behaviors in a \textsl{diverse} manner.
That is, we aim for an even distribution of exercised behaviors, rather than
focusing our testing effort solely on particular behaviors.
A simplified illustration of this idea can be found in Figure \ref{fig:motivation}.
In the figure, the grey areas in the input space represent the valid inputs (i.e., inputs that trigger the core functionality of the SUT),
whereas the white areas represent other invalid inputs (e.g., inputs that only trigger error handling code).
Traditional fuzzers like AFL~\cite{afl}, as depicted in Figure \ref{fig:motivation}.a, tend to produce malformed inputs and
hence only cover a small proportion of the valid behavior.
Further, the distribution of the triggered valid behavior mostly concentrates
on the more likely branches.
On the other hand, validity-focused fuzzers like Zest and RLCheck (cf. Figure \ref{fig:motivation}.b)
either only focus on covering many different behaviors
(but not their diverse execution, indicated by grey areas with a low concentration of green crosses)
or disproportionally test particular behaviors only (indicated by the small areas with a high number of green crosses).
Our proposed technique, \textsc{\ToolName} (cf. Figure \ref{fig:motivation}.c), avoids this problem
by searching for many different valid inputs while
evenly testing the behaviors in a diverse manner.

In particular, \textsc{\ToolName} distinguishes between \textsl{structure-changing} and
\textsl{structure-preserving} mutations.
Performing structure-changing mutations allows to search for new program behavior
that is only triggered if the input satisfies particular structural properties.
In contrast, the structure-preserving mutations allow to target specific behavior
of the code with different variants of the same input structure,
effectively testing the targeted behavior in a diverse manner.
\textsc{\ToolName} uses an adaptive mutation strategy that
utilizes the received program feedback
to guide the search towards high behavioral diversity.
In order to determine the behavioral diversity of a fuzzing campaign, 
we propose a novel metric that incorporates both the number of covered branches
and the branch execution distribution over all unique traces.

\vspace{10pt}
\noindent {\bf Novelty and Contributions.} Overall, we provide the following key contributions:
\begin{itemize}[leftmargin=0.5cm,topsep=0pt]
\item We present \textsc{\ToolName}, a novel fuzzing approach that generates valid and behaviorally diverse inputs.
\item We propose to utilize Hill numbers~\cite{Hill1973}, a common metric for species diversity from ecology, to quantify the behavioral diversity of the covered branches by a fuzzing campaign.
\item We evaluate \textsc{\ToolName} based on several fuzzing campaigns with XML and JavaScript SUTs, namely Ant, Maven, Closure, Rhino, Nashorn, and Tomcat.
\item We provide the source code of \textsc{\ToolName} and a replication package of our results at \url{https://github.com/hub-se/BeDivFuzz}. 
\end{itemize}

\vspace{10pt}
\noindent{\bf Significance of the Contributions.} We provide a novel metric that quantifies the \textsl{behavioral diversity} of a fuzzing campaign and propose a technique that improves upon the current state-of-the-art w.r.t. to that metric.
The proposed metric shifts the focus of simply \textsl{covering many behaviors} (i.e., branch coverage) to
\textsl{diversely testing many different behaviors}.
Thus, it will potentially serve as a stepping stone towards future systematic evaluations of a SUT with respect to reliability and correctness after the termination of a fuzzing campaign~\cite{BohmeLW21ResidualRisk,Boehme2018STADS}, as desired in practical software engineering.

\begin{figure}[t]
\centering
\begin{lstlisting}[frame=tlrb,tabsize=2, escapechar=|]
def generate_tree(depth=0):				|\label{treegen:line:1}|
	value = random.randint(0, 10)			|\label{treegen:line:2}|
	tree = BinaryTree(value)				|\label{treegen:line:3}|
	if depth >= MAX_DEPTH:				|\label{treegen:line:4}|
		return tree						|\label{treegen:line:5}|
	if random.choose([True, False]):		|\label{treegen:line:6}|
		tree.left = generate_tree(depth+1)	|\label{treegen:line:7}|
	if random.choose([True, False]):		|\label{treegen:line:8}|
		tree.right = generate_tree(depth+1)	|\label{treegen:line:9}|
	return tree							|\label{treegen:line:10}|
\end{lstlisting}
\caption{A simple binary tree generator. Adapted from \cite{reddy2020rlcheck}.}
\label{fig:tree_generator}
\end{figure}

\section{Background}\label{sec:background}
\subsection{Generator-Based Fuzzing}\label{subsec:generator_fuzzing}
Fuzzing (also known as fuzz testing) attempts to find bugs and crashes in software
through random input generation.
A common challenge in fuzzing is to produce test inputs for
programs that expect complex structured inputs (e.g., compilers).
Effectively testing these types of programs is challenging due to the following reasons:
(i) The generated inputs must be \textsl{syntactically} valid
in order to be successfully parsed by the program.
(ii) The input must also satisfy any additional \textsl{semantic} validity constraints
in order to actually exercise the SUT's core functionality.
(iii) The test inputs have to be \textsl{diverse} enough
to execute a variety of different program behavior.
Structure-aware fuzzers typically approach this problem by utilizing
domain-specific knowledge about the expected input type or format.
One way to provide this knowledge to the fuzzer is via a
declarative input specification (e.g., a grammar \cite{AschermannFHJST19NAUTILUS,eberlein2020egbf, Godefroid2008GrammarWhiteBoxFuzzing,Holler2012FuzzingCodeFragments, NguyenNKG20MoFuzz, soremekun2020inputs, Wang2019Superion}).
Alternatively, the tester may write an imperative \textsl{generator} program that
randomly samples syntactically valid inputs, an approach known as
\textsl{generator-based fuzzing} \cite{junit-quickcheck, PadhyeLS19JQF, Padhye2019Zest}.
For example, Figure~\ref{fig:tree_generator} shows the pseudocode for a generator that
produces random binary trees,
which will serve as the running example throughout the paper.
The function \texttt{generate\_tree} first samples a random integer value between 0 and 10 (Line~\ref{treegen:line:2})
to instantiate the root node (Line~\ref{treegen:line:3}).
If a user-defined maximum depth has been reached,
the generated node is returned (Lines~\ref{treegen:line:4}--\ref{treegen:line:5}).
Otherwise, the generator non-deterministically decides whether to
generate left and right child nodes, in which case
 \texttt{generate\_tree} is recursively called (Lines~\ref{treegen:line:6}--\ref{treegen:line:9}).
The output of the generator thus depends on
the sequence of \textsl{random choices} made during the generation process.
In the example, the sequence of random choices precisely determines
the shape of the binary tree and all the tree node values.
As a concrete example, consider the following random choice sequence
for the call of \texttt{generate\_tree} (\texttt{MAX\_DEPTH = 5}):

\vspace{-10pt}
\begin{table}[h!]
  \begin{center}
    \begin{tabular}{ll}
      \textbf{Random choice $\rightarrow$ result} & \textbf{Context}\\
      \hline
      \texttt{random.randint} $\rightarrow$ \texttt{3}  & Root: node value  (Line~\ref{treegen:line:2})\\
      \texttt{random.choose} $\rightarrow$ \texttt{False}  & Root: generate left child?  (Line~\ref{treegen:line:6})\\
      \texttt{random.choose} $\rightarrow$ \texttt{False}  & Root: generate right child?  (Line~\ref{treegen:line:8})\\
    \end{tabular}
  \end{center}
\end{table}

\noindent This sequence of random choices would result in a ''binary-tree'' consisting of
a single root node with the value 3 and no child nodes:

\begin{center}
	\begin{tikzpicture}[
  	every node/.style = {minimum width = 1em, draw, circle},
  	]
  	\node {3};
\end{tikzpicture}
\end{center}

\noindent To give a different example, consider the following choice sequence:

\vspace{-10pt}
\begin{table}[h!]
  \begin{center}
    \begin{tabular}{ll}
      \textbf{Random choice $\rightarrow$ result} & \textbf{Context}\\
      \hline
      \texttt{random.randint} $\rightarrow$ \texttt{3}  & Root: node value  (Line~\ref{treegen:line:2})\\
      \texttt{random.choose} $\rightarrow$ \texttt{True}  & Root: generate left child?  (Line~\ref{treegen:line:6})\\
      \texttt{random.randint} $\rightarrow$ \texttt{5}  & Left child: node value  (Line~\ref{treegen:line:2})\\
      \texttt{random.choose} $\rightarrow$ \texttt{False}  & Left child: gen. left child?  (Line~\ref{treegen:line:6})\\
      \texttt{random.choose} $\rightarrow$ \texttt{False}  & Left child: gen. right child?  (Line~\ref{treegen:line:8})\\
      \texttt{random.choose} $\rightarrow$ \texttt{True}  & Root: generate right child?  (Line~\ref{treegen:line:8})\\
      \texttt{random.randint} $\rightarrow$ \texttt{7}  & Right child: node value  (Line~\ref{treegen:line:2})\\
      \texttt{random.choose} $\rightarrow$ \texttt{False}  & Right child: gen. left child?  (Line~\ref{treegen:line:6})\\
      \texttt{random.choose} $\rightarrow$ \texttt{False}  & Right child: gen. right child?  (Line~\ref{treegen:line:8})\\
    \end{tabular}
  \end{center}
\end{table}

\noindent The corresponding binary tree consists of a root node with the value 3 and
leaf nodes with the values 5 and 7 as the left and right child, respectively:

\begin{center}
	\begin{tikzpicture}[
  	every node/.style = {minimum width = 1em, draw, circle},
  	]
  	\node {3}
  	child {node {5}}
  	child {node {7}};
\end{tikzpicture}
\end{center}

A typical challenge in generator-based testing arises when the input is expected to satisfy complex semantic validity constraints
that are not explicitly considered by the generator.
To continue the current example, the SUT may expect
the generated tree to be sorted (e.g., a binary search tree) or height-balanced (e.g., an AVL tree).
In this case, random input sampling is usually not very effective as
most of the generated inputs will be rejected by the SUT.
Instead, recent techniques bias the generator towards producing valid inputs by
directly controlling the random choices made during the input generation process.

Zest~\cite{Padhye2019Zest} relies on the fact that
the implementation of random number generators usually
depends on a pseudo-random source of \textsl{untyped bits},
which Padhye et al. refer to as \textsl{parameters}.
Specifically, random values of a particular type are generated by
consuming and interpreting a fixed number of parameters.
For instance, a random boolean value may be sampled by
consuming one bit from the source of untyped parameters, which is then
interpreted as \texttt{False} if it is zero and otherwise as \texttt{True}.
The pseudo-random nature of this method ensures that
using the same random seed always generates the same sequence of parameters.
For a given generator, this will produce the same sequence of random choices,
effectively resulting in the same generated input.
On the other hand, by mutating the untyped parameter sequence,
the sequence of random choices can be altered, which
corresponds to complex structural mutations in the input domain.
For example, mutating the parameters that are consumed by the first call to
\texttt{random.randint()} (Figure \ref{fig:tree_generator}, Line~\ref{treegen:line:2})
results in the value of the root node to be mutated.
Similarly, mutating the parameters corresponding to a call to
\texttt{random.choose()} (Line~\ref{treegen:line:6}) may
change the decision on whether a left child node will be generated.
As a result, by controlling the sequence of parameters,
it is possible to directly control the output of the input generator.
Zest \cite{Padhye2019Zest} exploits this insight by performing a feedback-directed parameter search to
guide the generators towards high coverage in the semantic analysis stages.
Contrary, RLCheck \cite{reddy2020rlcheck} employs reinforcement learning to
automatically learn a \textsl{guide} that
leads the generator to produce diverse valid inputs (w.r.t. a validity function).

\subsection{Hill-Numbers} \label{subsec:hill_numbers}
In the field of ecology, researchers are interested in quantifying the biodiversity 
of an assemblage based on a sample of species.
A commonly used class of diversity metrics are the Hill numbers~\cite{Hill1973}.
The Hill numbers consider both species richness (i.e., the total number of different species) and
species abundances (i.e., the number of individuals per species) in a sample,
and are defined as follows:

\begin{definition}[Hill number of order $q$]
	Let $S$ be the species richness and $p_i$ the
	relative abundance of the $i$-th species in the dataset. The \textsl{Hill number of order $q$} is defined as ($^qD$ is the original notation):
	\begin{equation} \label{eq:hill_numbers}
		^qD = D(q) = \left(\sum^S_{i=1} p_i^q\right)^{1/(1-q)}
	\end{equation}
\end{definition}

\noindent For $q = 1$, Equation~\ref{eq:hill_numbers} is undefined,
but its limit for $q\rightarrow 1$ corresponds to
the exponential of the Shannon(-Wiener) index \cite{Shannon48,SpellerbergF03ShannonWienerIndex}:

\begin{equation} \label{eq:hill_number_q1}
	^1D = D(1) = \lim_{q \rightarrow 1} {^qD} = \exp \left(-\sum^S_{i=1} p_i \log p_i \right)
\end{equation}

The order $q$ determines the sensitivity of the metric to the relative species abundances,
where typically measures for $q \in \{0, 1, 2\}$ are reported.
For $q = 0$, the relative species abundances are not considered at all and
$D(0)$ corresponds to the total number of species in the dataset (i.e., the richness).
For $q = 1$, the species are weighted in proportion to their relative abundances,
which is why $D(1)$ can be interpreted as the effective number of ''typical'' species.
For $q = 2$, more weight is given to the most abundant species and the rarer species are discounted.
As a result, $D(2)$ can be interpreted as the effective number of ''common'' or dominant species.

\newcommand{\choiceSequence}{\gamma}
\newcommand{\paramSequence}{\sigma}

\section{Approach} \label{sec:approach}
In this section, we first describe our approach to 
extend generator-based fuzzing
with the ability to produce more behaviorally diverse inputs (Section \ref{subsec:overview}--\ref{subsec:fuzzing_algorithm}).
Then, we introduce a novel metric to quantify 
the behavioral diversity of a fuzzing campaign based on 
Hill numbers (Section \ref{subsec:behavioral_diversity}).

\subsection{Overview}  \label{subsec:overview}
The key idea of our approach can be summarized as follows: 
\begin{enumerate}
\item We search for interesting \textsl{input structures} in the space of valid inputs through
\textsl{structure-changing} mutations.
\item We produce different variants of the same input structure by applying
\textsl{structure-preserving} mutations with the goal of
exploring diverse execution traces.
\end{enumerate}

To bias the input generation towards high behavioral diversity,
we observe the coverage and validity feedback from the program after every execution
and adapt the mutation strategy accordingly.

\subsection{Structural Parameter Splitting}\label{subsec:parameter_splitting}
As described in Section \ref{subsec:generator_fuzzing}, the non-determinism of
the input generation process is entirely controlled by the sequence of random choices made.
Our first insight is that the choices can be
classified into two different types:
\textsl{structural choices} and \textsl{value choices},
depending on their influence on the control-flow behavior of the generator.
Based on the notion of a \textsl{choice point} by Reddy et al. \cite{reddy2020rlcheck},
we define these choice types as follows:

\begin{definition}[Structural and Value choices]
	Let the choice point $p$ be a tuple $(\ell, C)$,
	where $\ell \in \mathbb{L}$ is a program location in the generator $G$ and
	$C \subseteq \mathbb{C}$ is a finite domain of choices.
	The choice point $p$ is said to produce \textsl{structural choices} if
	the evaluation of a branch condition at
	some point during the execution of $G$ depends on the choice $c \in C$.
	Otherwise, $p$ is said to produce \textsl{value choices}.
\end{definition}

For example, the choice point $p_1 = $~(Line \ref{treegen:line:2}, \texttt{[0, 1, ..., 10]}) in Figure \ref{fig:tree_generator} produces
value choices, since no branch condition in \texttt{generate\_tree} depends on that choice.
On the other hand, both choice points $p_2 = $~(Line \ref{treegen:line:6}, \texttt{[True, False]}) and
$p_3 = $~(Line \ref{treegen:line:8}, \texttt{[True, False]}) produce structural choices,
since the corresponding choices directly influence the branching behavior of the generator $G$
(i.e., the decision on whether to generate child nodes or not).

Based on this insight, our first idea is to \textsl{split} the sequence of untyped parameters
into two distinct parameter sequences based on the choice type they are used for:

\begin{enumerate}
	\item A \textsl{structural parameter sequence} to be consumed by structural choice points.
	\item A \textsl{value parameter sequence} to be consumed by value choice points.
\end{enumerate}
That is, we can now represent each input as a tuple $(\sigma_s, \sigma_v)$,
where $\sigma_s$ and $\sigma_v$ represent the structural and value parameters consumed by the input generator, respectively.
To further illustrate this idea, consider the following binary tree:
\begin{center}
	\begin{tikzpicture}[
  	every node/.style = {minimum width = 1em, draw, circle},
  	]
  	\node {3}
  	child {node {5}}
  	child {node {7}};
\end{tikzpicture}
\end{center}

The corresponding parameter sequence $\sigma$ (shown as typed choice values for the sake of simplicity) is given by:
\begin{align*}
	\sigma =3, \texttt{True}, 5, \texttt{False}, \texttt{False}, \texttt{True}, 7, \texttt{False}, \texttt{False}
\end{align*}

Here, the structural choices consists of all the boolean and the value choices of all the integer values.
Therefore, we can represent this parameter sequence by the following two distinct sequences:
\begin{align*}
\sigma_s &= \texttt{True}, \texttt{False}, \texttt{False}, \texttt{True}, \texttt{False}, \texttt{False} \\
\sigma_v &= 3, 5, 7
\end{align*}

This change in how the parameters for a generator are handled is a key idea of our approach,
since it gives us access to the following operations and concepts.

First, it enables us to perform more controlled mutations on the input.
In particular, by mutating the structural parameter sequence,
we can directly mutate the structural choices made during the input generation process.
By definition, the change of structural choices eventually results in
a change of the control-flow behavior in the generator, which typically
produces an input with a different structure.
For example, if we consider the binary tree generator from Figure \ref{fig:tree_generator},
mutating the structural parameters changes the boolean decisions
on the generation of child nodes.
On the other hand, by mutating the value parameters only,
the control-flow behavior of the generator is preserved,
yet different value choices are sampled.
In the case of the binary tree generator, mutating value parameters results in
mutated choices for the node values,
while keeping the shape of the binary tree unmodified.
Overall, the access to these structure-changing and structure-preserving
mutations allows us to explore the input space in a more controlled manner.

The second benefit of separating the structural and value parameters is that
it allows us to synthesize an \textsl{abstract input} from a concrete input:

\begin{definition}[Abstract Input, Structural Signature]
	Let $i$ be an input generated by a generator $G$,
	$\sigma_s = c_{s_1}, c_{s_2}, \dots, c_{s_n}$ be the corresponding
	sequence of structural parameters, and
	$\sigma_v = c_{v_1}, c_{v_2}, \dots, c_{v_m}$ be the corresponding
	sequence of value parameters made by $G$ during the generation of $i$.
	The \textsl{abstract input} $A(i)$ of the input $i$ is the input
	that is obtained by setting all value parameters in $\sigma_v$ as unspecified (or symbolic),
	while fixing the concrete structural parameters $\sigma_s$.
	The sequence of concrete structural parameters $\sigma_s$ is called the
	\textsl{structural signature} of $A(i)$.
\end{definition}

Intuitively, the abstract input represents the structural skeleton of the input,
which may be concretized by specifying the missing value parameters.
In our running example, the abstract input of a generated binary tree
would correspond to a binary tree with the same shape but unspecified node values.
As an example, for the above described tree,
the abstract input $A(t)$ fixes the structural parameters $\sigma_s$ and leaves the value parameters $\sigma_v$ unspecified ($c_{v_1}, c_{v_2}, c_{v_3}$), i.e.:
\begin{center}
	\begin{tikzpicture}[
  	every node/.style = {minimum width = 1em, draw, circle},
  	]
  	\node {$c_{v_1}$}
  	child {node {$c_{v_2}$}}
  	child {node {$c_{v_3}$}};
\end{tikzpicture}
\begin{align*}
	\sigma_s &= \texttt{True}, \texttt{False}, \texttt{False}, \texttt{True}, \texttt{False}, \texttt{False}  \\
	\sigma_v &= c_{v_1}, c_{v_2}, c_{v_3}
\end{align*}
\end{center}

\noindent The structural signature (i.e., the sequence of structural parameters $\sigma_s$) of $A(t)$ thus describes
the set of all binary trees consisting of three nodes:
a root node that is connected to two child nodes where each node has a value between 0 and 10
(the domain of the choice point $p_1$ = (Line \ref{treegen:line:2}, \texttt{[0, 1, ..., 10]})).
Since we represent each input as a tuple $(\sigma_s, \sigma_v)$,
we can easily check whether two inputs share the same abstract input
by comparing their structural signatures.
This is another crucial aspect of our approach,
as this property allows us to reason about the explored input space
on a higher level.
In particular, our approach uses the concept of abstract inputs in order to
identify interesting input structures that exercise new behavior.

\subsection{Feedback-Driven Search Strategy} \label{subsec:fuzzing_algorithm}
The fuzzing algorithm of \textsc{\ToolName} is presented in Algorithm \ref{alg:fuzzing_algorithm}.
It is based on the coverage-guided Zest algorithm for generator-based fuzzing,
and is extended by integrating the concepts of
structural parameter splitting and abstract input structures.
In particular, the two key components of the \textsc{\ToolName} algorithm consist of:
\begin{enumerate}
	\item An adaptive mutation strategy that biases input generation
towards high behavioral diversity, and
\item A fuzzing heuristic that guides the search strategy based on
the structural properties of the input.
\end{enumerate}
We highlight these main innovations in Algorithm \ref{alg:fuzzing_algorithm} in grey.

\begin{algorithm}[htp]
  \SetAlgoLined\DontPrintSemicolon
  \footnotesize
	\caption{\textsc{\ToolName} Algorithm}
	\label{alg:fuzzing_algorithm}
	\SetKwInOut{Input}{Input}%
	\SetKwInOut{Output}{Output}%
	\Input{
		program $p$, generator $g$ \\
	}
	\Output{
		a set of test inputs, a set of failing inputs	
	}
	\BlankLine
  	$Q \gets \{\textsc{random}\}$\tcc*{Initial seed input} \label{alg:fuzz:init_queue}
	$F \gets \emptyset$\tcc*{Failing parameters} \label{alg:fuzz:init_failures}
	$C \gets \emptyset$\tcc*{Total coverage} \label{alg:fuzz:init_coverage}
	\HiLi $T \gets \emptyset$\tcc*{Unique traces} \label{alg:fuzz:init_traces}
	\HiLi $S \gets \emptyset$\tcc*{Interesting valid input structures} \label{alg:fuzz:init_structures}
	\Repeat{given time budget expires}{ \label{alg:fuzz:begin_fuzzing_loop}
		\For(){$(\sigma_s, \sigma_v)$ in $Q$}{ \label{alg:fuzz:select_input}
			\For(){$1 \leq i \leq$ \Call{numChildren}{$\sigma_s, \sigma_v$}}{ \label{alg:fuzz:determine_numChildren}
				\HiLi $\tilde{\sigma}_s, \tilde{\sigma}_v \gets \Call{mutateAdaptive}{\sigma_s, \sigma_v}$\; \label{alg:fuzz:mutate_params}
				$\mathit{input} \gets g(\sigma_s, \sigma_v)$\; \label{alg:fuzz:generate_input}
				$\mathit{coverage}, \mathit{result} \gets \Call{run}{p, \mathit{input}}$\; \label{alg:fuzz:run_input}
				\If(\tcc*[f]{Save failures}){$\mathit{result} = \textsc{Failure}$}{ \label{alg:fuzz:check_failure}
					$F \gets F \cup \{(\sigma_s, \sigma_v)\}$\; \label{alg:fuzz:add_failure}
				}
				\Else(\tcc*[f]{Check if input has interesting structure}){
					\HiLi \If{$\mathit{result} = \textsc{Valid}$ and $\mathit{coverage} \nsubseteq C$ and $\sigma_s \notin S$}{ \label{alg:fuzz:check_toSave}
						$Q \gets Q \cup \{(\sigma_s, \sigma_v)\}$ \tcc*{Add to queue} \label{alg:fuzz:save_input}
						\HiLi $S \gets S \cup \{\sigma_s\}$ \tcc*{Add saved structure} \label{alg:fuzz:save_structure}
					}
					\Call{updateCoverageStats}{$\mathit{coverage}, \mathit{result}, C, T$} \label{alg:fuzz:update_coverage}
				}
			}
		}
	} \label{alg:fuzz:end_fuzzing_loop}
  \KwRet $g(Q), g(F)$\;
\end{algorithm}

Similar to Zest (and other coverage-guided fuzzers),
we maintain a queue $Q$ of interesting inputs,
which is initially seeded with a random value (Line~\ref{alg:fuzz:init_queue}).
While Zest operates on single parameter sequences $\sigma$,
\textsc{\ToolName} operates on tuples $(\sigma_s, \sigma_v)$ of split parameter sequences.
For the sake of brevity, in the remainder of this paper we will use the term \textsl{parameters} when
referring to tuples $(\sigma_s, \sigma_v)$ of split parameter sequences.
Throughout the fuzzing campaign, the algorithm maintains
basic bookkeeping data, such as the set of failing parameters and the current branch coverage
(Lines~\ref{alg:fuzz:init_failures}--\ref{alg:fuzz:init_coverage}).
Additionally, \textsc{\ToolName} keeps track of
all unique coverage traces (Line~\ref{alg:fuzz:init_traces})
and the set of interesting abstract input structures (Line~\ref{alg:fuzz:init_structures}).
The set of unique coverage traces is used to adaptively bias the mutation strategy,
described in more detail at the end of this section.
The set of abstract input structures is utilized by \textsc{\ToolName}
to guide the search within the space of valid inputs.

The main fuzzing loop~(Lines \ref{alg:fuzz:begin_fuzzing_loop}--\ref{alg:fuzz:end_fuzzing_loop})
has the same structure as other coverage-guided fuzzers, described as follows:
First, a parameter sequence is selected from the queue (Line~\ref{alg:fuzz:select_input}),
after which the number of child parameters is determined (Line~\ref{alg:fuzz:determine_numChildren}).
We use the same heuristic as Zest, which computes the number of child parameters
based on the coverage of the parent input.
Each child is generated by performing one or more mutations
on the parent parameter sequence (Line~\ref{alg:fuzz:mutate_params}).
The mutated parameters are then used to
generate a concrete input using the provided generator (Line~\ref{alg:fuzz:generate_input}).
Afterwards, the system under test is executed with the generated input, which
yields the code coverage and the validity feedback (Line~\ref{alg:fuzz:run_input}).
The validity feedback is stored in the variable $\mathit{result}$ and
can be any of $\{\textsc{Valid}, \textsc{Invalid}, \textsc{Failure}\}$.
In the case of $\textsc{Failure}$, the current parameters are saved to
the set of failures (Line~\ref{alg:fuzz:add_failure}).
Otherwise, the algorithm heuristically decides
whether the current parameters should be saved to the queue
(i.e., if the parameters are interesting enough to be further mutated)
based on the observed execution results (Line~\ref{alg:fuzz:check_toSave}).
This is by default the case if the input is (i) valid, (ii) exercises new coverage, and
(iii) represents an input structure that has not
been previously added to the queue.
Finally, the bookkeeping-data is updated (Line~\ref{alg:fuzz:update_coverage}).
This includes updating the branch coverage and the set of unique coverage traces.

As described in Section \ref{subsec:parameter_splitting},
\textsc{\ToolName} is able to distinguish between \textsl{structure-changing} and
\textsl{structure-preserving} mutations.
This is done by performing mutations on either the structural parameters $\sigma_s$
or on the value parameters $\sigma_v$.
Performing structure-changing mutations allows to search for new program behavior
that is only triggered if the input satisfies particular structural properties.
In contrast, structure-preserving mutations allow to test a specific behavior
of the code with different variants of the input
(i.e., inputs with the same abstract structure but different values).

In order to decide which type of mutation to perform, \textsc{\ToolName}
keeps track of the types of mutations that have been performed on a parameter sequence
and observes whether the resulting input has resulted in the execution of a \textsl{unique trace}.
The following mutations are then biased towards the mutation operator that
is more likely to produce unique execution traces, as illustrated in Algorithm \ref{alg:mutation}.
Specifically, with a probability of $\epsilon$, a random type of mutation is performed.
Otherwise, the mutation type is chosen based on heuristic scores $R_s$ and $R_v$
for the structural and value mutations, respectively (Line~\ref{alg:mutation:get_rewards}).
For a given mutation type $x$, the score is calculated as the fraction of inputs that have exercised a unique trace ($U_x$) out
of all the inputs that were generated by the respective mutation type ($N_x$), i.e.:
\begin{equation} \label{eq:heuristic_score}
		R_x = \frac{U_x}{N_x} \quad x \in \{s, v\}
\end{equation}
Thus, we bias the selection of the mutation operator towards producing inputs
that diversely exercise particular behaviors,
while relying on the structural search heuristic in Algorithm~\ref{alg:fuzzing_algorithm} (Line~\ref{alg:fuzz:check_toSave})
to discover new behaviors.

\begin{algorithm}[htpb]
	\footnotesize
	\caption{Adaptive Mutation Strategy}
	\label{alg:mutation}
	\SetKwInOut{Input}{Input}
	\SetKwInOut{Output}{Output}
	\SetKw{KwGoTo}{goto}
	\Input{
		structural parameters $\sigma_s$, value parameters $\sigma_v$, exploration factor $\epsilon$
	}
	\Output{
		mutated parameters $\tilde{\sigma}_s, \tilde{\sigma}_v$
	}
	\BlankLine
	\If{$\Call{uniformRandom}{\null} < \epsilon$}{ \label{alg:mutation:random_mutation}
		\If(\tcc*[f]{Mutate either of the params}){\Call{randomBoolean}{\null}}{\label{alg:mutation:random_mutation}
			\KwRet $(\Call{mutate}{\sigma_s}, \sigma_v)$
		}
		\Else{
			\KwRet $(\sigma_s, \Call{mutate}{\sigma_v})$
		}
	}
		
	$R_s, R_v \gets \Call{calculateScores}{\null} $ \tcc*[f]{Score by mutation type (Eq. 4)} \label{alg:mutation:get_rewards}
	
	\If(\tcc*[f]{Perform most promising mutation}){$R_s \neq R_v$}{ \label{alg:mutation:begin_choose_mutation}
		\If{$R_s > R_v$}{
			\KwRet $(\Call{mutate}{\sigma_s}, \sigma_v)$ \tcc*[f]{Mutate structural params}
		}
		\Else{
			\KwRet $(\sigma_s, \Call{mutate}{\sigma_v})$ \tcc*[f]{Mutate value params}
		}
	} \label{alg:mutation:end_choose_mutation}
	\Else{
		\KwGoTo \ref{alg:mutation:random_mutation} \tcc*[f]{Resort to random mutation} \label{alg:mutation:fallback_random_mutation}
	}
\end{algorithm}

\subsection{A Metric for Behavioral Diversity} \label{subsec:behavioral_diversity}
As part of this work, we are interested in establishing a notion of \textsl{behavioral diversity}
in the context of fuzz testing.
In particular, we propose to use \textsl{Hill numbers} (Section~\ref{subsec:hill_numbers}), or \textsl{effective number of species} \cite{Hill1973}.
The inspiration for this metric stems from
the STADS (\textsl{Software Testing and Analysis as Discovery of Species}) framework introduced by Böhme~\cite{Boehme2018STADS}.
In this framework, Böhme links the sampling process in an assemblage for species discovery to
the process of sampling from the program's input space to discover particular features of the input's execution
(e.g., covered branches, reached program states).
However, rather than using the framework to estimate species discovery probabilities~\cite{Boehme2018STADS,BohmeLW21ResidualRisk},
we build upon the same connection between ecology and software testing by applying an established biodiversity index
--- known as Hill numbers --- to quantify the behavioral diversity of a fuzzing campaign.

We apply this diversity measure to the context of fuzzing as follows.
Consider a fuzzing campaign, where a fuzzer $F$ samples inputs $I$ to fuzz a program $P$.
Executing $P$ with an input $i \in I$ results in a trace $t(i)$ that consists of
the sequence of branches visited during the execution.
Each branch $b$ that has been covered by at least one input $i$ can now be seen as a species.
The abundance of a species (i.e., of a covered branch $b$) can then be computed as the number $c(b)$
of unique traces that have executed the branch $b$ (the branch execution count).
Given these measures, we can now quantify the \textsl{behavioral diversity} of a fuzzing campaign as
the Hill numbers computed over the distribution of branch execution counts.

\begin{definition}[Behavioral Diversity]
	Let $I$ be the set of inputs generated by a fuzzer during a fuzzing campaign, and
	let $C: \mathcal{B} \mapsto \mathbb{N}_{>0} $ be a function that maps a covered branch $b \in \mathcal{B}$ to its relative execution count over all unique traces.
	Then, the \textsl{behavioral diversity} of order $q$ is defined as:
	\begin{equation} \label{eq:behavioral_diversity}
		B(q) = \left(\sum^{}_{b \in \mathcal{B}} C(b)^q\right)^{1/(1-q)}
	\end{equation}
\end{definition}

Intuitively, this measure quantifies the \textsl{effective number} of
diversely covered branches,
that is, the number of branches that were equally often executed by diverse inputs (i.e., inputs with unique traces).
While $B(0)$ is equal to the number of covered branches (since it equally weights rare and common branches),
$B(1)$ weights branches by their relative execution counts and can thus be interpreted as
the ''effective number of typically covered branches''.
$B(2)$ gives more weight to common branches and
can be seen as the ''effective number of common branches''.

In the context of fuzzing, behavioral diversity considers 
the distribution of diverse branch execution counts in order to measure the 
degree of exploration bias towards specific program behavior. 
Higher numbers indicate that a fuzzer is diversely exploring more branches, 
whereas low numbers mean that the fuzzer is spending most of the 
resources exploring the same few behaviors. 
Low behavioral diversity is most evident for random blackbox-techniques, 
which may be able to spuriously cover many branches, 
but ultimately exercise the same likely code regions over and over again. 
Especially for “unsuccessful” fuzz campaigns where no bugs are found, 
behavioral diversity can provide valuable insights into the progress of a fuzzer 
as a complementary metric to branch coverage.
For instance, practitioners may consider more diversely executed branches as 
more reliable than less diversely executed ones, since they have been tested more “thoroughly”. 
On the other hand, researchers may tune their fuzzers to repeatedly target less diversely 
executed branches in order to increase the overall confidence in the 
reliability of the respective behavior.

\section{Evaluation}
In this section, we evaluate the effectiveness of \textsc{\ToolName} in generating inputs
that exhibit diverse behavior.
We compare our approach against three techniques:
RLCheck~\cite{reddy2020rlcheck}, Zest \cite{Padhye2019Zest}, and an implementation of Quickcheck~\cite{ClaessenH00Quickcheck} from the JQF~\cite{PadhyeLS19JQF} fuzzing framework.
In particular, we seek to answer the following research questions:
\begin{itemize}
 \item[RQ1] Is \textsc{\ToolName} able to effectively produce diverse valid inputs for real-world benchmarks? (Section~\ref{sec:results-RQ1}) \label{sec:evaluation:RQ1}

 \item[RQ2] Do the inputs generated by \textsc{\ToolName} have a higher behavioral diversity compared to state-of-the-art techniques? (Section~\ref{sec:results-RQ2}) \label{sec:evaluation:RQ2}

 \item[RQ3] How does \textsc{\ToolName} perform in terms of fault finding capabilities? (Section~\ref{sec:results-RQ3}) \label{sec:evaluation:RQ3}
\end{itemize}

\subsection{Study Design}

\subsubsection*{Baseline Techniques}
We compare against two state-of-the-art techniques in generator-based testing, namely
RLCheck~\cite{reddy2020rlcheck} and Zest~\cite{Padhye2019Zest}.
RLCheck uses reinforcement learning in order to automatically learn a \textsl{guide} that
leads the generator towards producing many diverse valid inputs.
Zest is a generator-based fuzzer that integrates validity feedback into
a coverage-guided search algorithm to effectively cover the semantic analysis code.
We also compare against a Java version of Quickcheck \cite{ClaessenH00Quickcheck}
by running Zest without code coverage and validity feedback,
as done in the original evaluation of RLCheck.
That is, Quickcheck performs random sampling of inputs using the generator.

\subsubsection*{Experimental Subjects}
Our evaluation is conducted on six real-world benchmarks,
namely Apache Ant, Apache Maven, Mozilla Rhino,
Google Closure Compiler, Oracle Nashorn, and Apache Tomcat.
The first four subjects have been used in
the original evaluations of Zest~\cite{Padhye2019Zest} and RLCheck~\cite{reddy2020rlcheck};
we add two additional subjects for a broader benchmark.
In addition, we have updated the subjects to the latest versions
available at the time when we conducted the experiments.
Inputs for Ant, Maven, and Tomcat are generated by an
XML generator, whereas Rhino, Closure, and Nashorn
use a JavaScript code generator.

\subsubsection*{Configuration}
We run all baseline techniques with their default configurations.
In order to run Nashorn and Tomcat with RLCheck, we use configuration files based on
those provided for Rhino and Maven by the original authors, respectively.
For \textsc{\ToolName}, we use $\epsilon = 0.2$ as the probability
to perform a random mutation.
Further, we provide results for two configurations of \textsc{\ToolName},
which we refer to as \textsc{\ToolName-structure} and \textsc{\ToolName-simple}.
\textsc{\ToolName-structure} utilizes both structural mutations and analysis of input structure novelty,
whereas \textsc{\ToolName-simple} only employs structural mutations.
In order to make the input generators used by Zest compatible with our technique,
we manually extended them to handle split parameter sequences
and annotated the choice points with the type of choice they produce.
However, we note that the latter could also be done automatically
through data-flow analysis.

\subsubsection*{Implementation}
We have implemented \textsc{\ToolName}~\cite{BeDivFuzzTool} as an extension of Zest in JQF~\cite{PadhyeLS19JQF}.
JQF is a framework for generator-based testing in Java and
implements different fuzzing algorithms, including the
Zest algorithm and AFL-like fuzzing for Java programs.

\subsubsection*{Experimental Parameters}
For the evaluation of RQ1 (input diversity) and RQ2 (behavioral diversity),
we run each experiment with a timeout of 1 hour.
The original evaluation of RLCheck~\cite{reddy2020rlcheck} was conducted with a timeout of 5 minutes,
as the authors assumed a property-based testing context where 
the generator is typically executed only for a short time.
While a 5 minute timeout may seem short compared to common timeouts in fuzzing evaluations, 
it should also be noted that collecting coverage metrics for blackbox methods (e.g., Quickcheck or RLCheck) is significantly more expensive
than for greybox methods (e.g., Zest or \textsc{\ToolName}).
In particular, the inputs for blackbox techniques first need to be generated on the uninstrumented SUT until timeout, 
and subsequently be replayed on the instrumented version of the SUT.
This is necessary to allow for a fair comparison against greybox methods, since instrumentation adds additional overhead.
However, this methodology results in significantly longer experimental runtimes\footnote{In our experiments, we have observed up to $50\times$ longer runtimes when computing coverage for Quickcheck and RLCheck.}.
On the other hand, coverage data may not be meaningful if timeouts are too short, 
since coverage-guided algorithms typically need some time to become effective.
Thus, for the evaluation of RQ1 (input diversity) and RQ2 (behavioral diversity),
we extend the timeout for each experiment to 1 hour.
We further justify this decision by our observation that 
coverage has plateaued in almost all experiments after this timeout.
However, for RQ3 (fault finding capabilities),
we extend the timeout to 24 hours, since
evaluations that focus on bug metrics should 
be performed with longer timeouts, 
as suggested by Klees et al. \cite{Klees2018EvaluateFuzzing}.
To account for the variability of
the results due to randomness, we perform 30 repetitions. 
Statistical significance is assessed using
the Mann-Whitney $U$ test (also known as the Wilcoxon rank-sum test) with $\alpha = 0.01$,
as suggested by Arcuri and Briand \cite{Arcuri2014AssessFuzzing} for randomized algorithms.

As for hardware, we conducted all experiments
on a server with an Intel(R) Xeon(R) E7-4880 2.5GHz CPU and
1TB of RAM running openSUSE Leap 15.

\subsection{RQ1: Generating Diverse Valid Inputs} \label{sec:results-RQ1}

\newcommand{\divFigureWidth}{1.55in}
\begin{figure}[htpb]
\begin{tabular}{c}
	\includegraphics[width=3in]{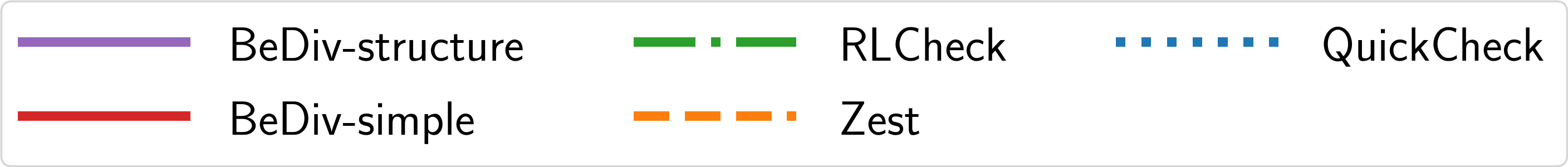} \\[-2ex]
	\captionsetup[subfloat]{captionskip=-6pt}
	\subfloat[Ant]{
		\includegraphics[width = \divFigureWidth]{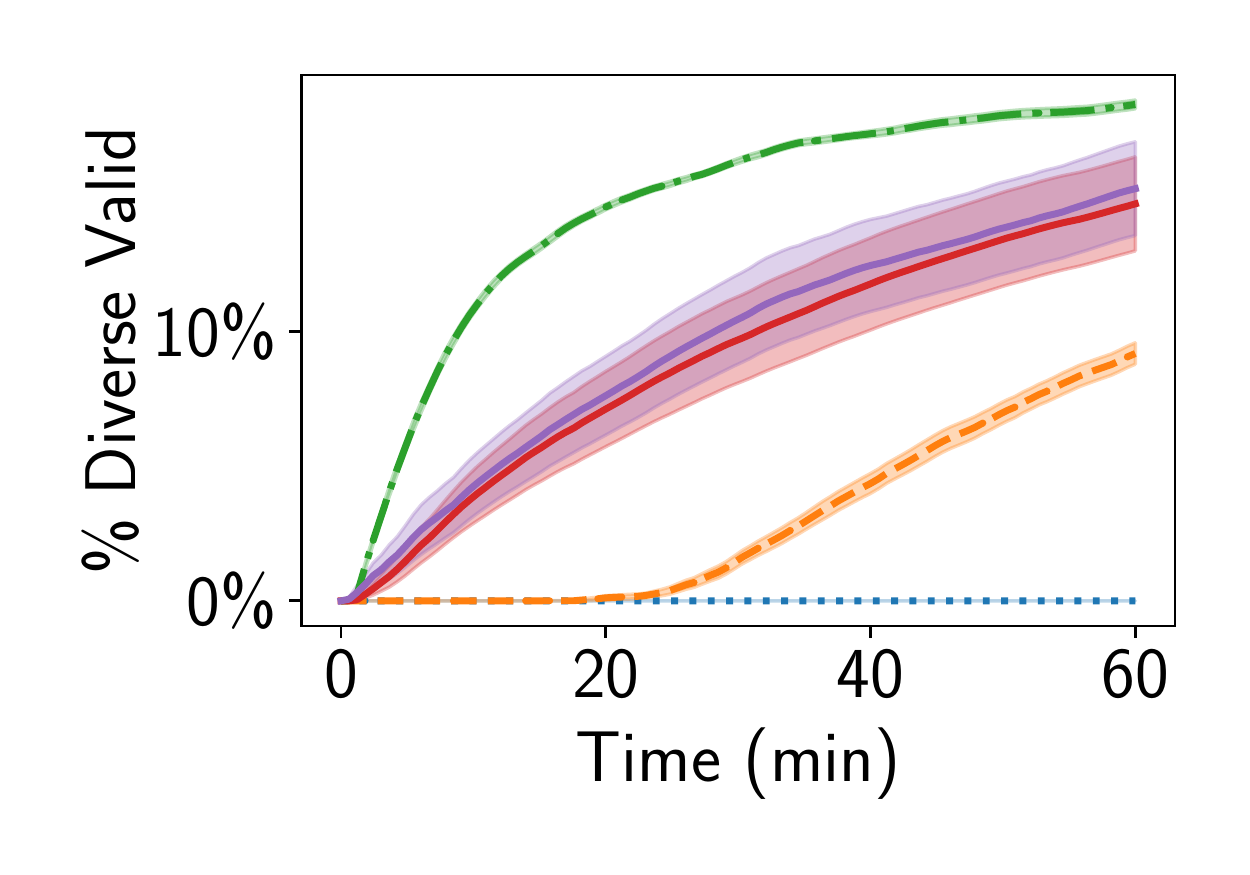}
		\includegraphics[width = \divFigureWidth]{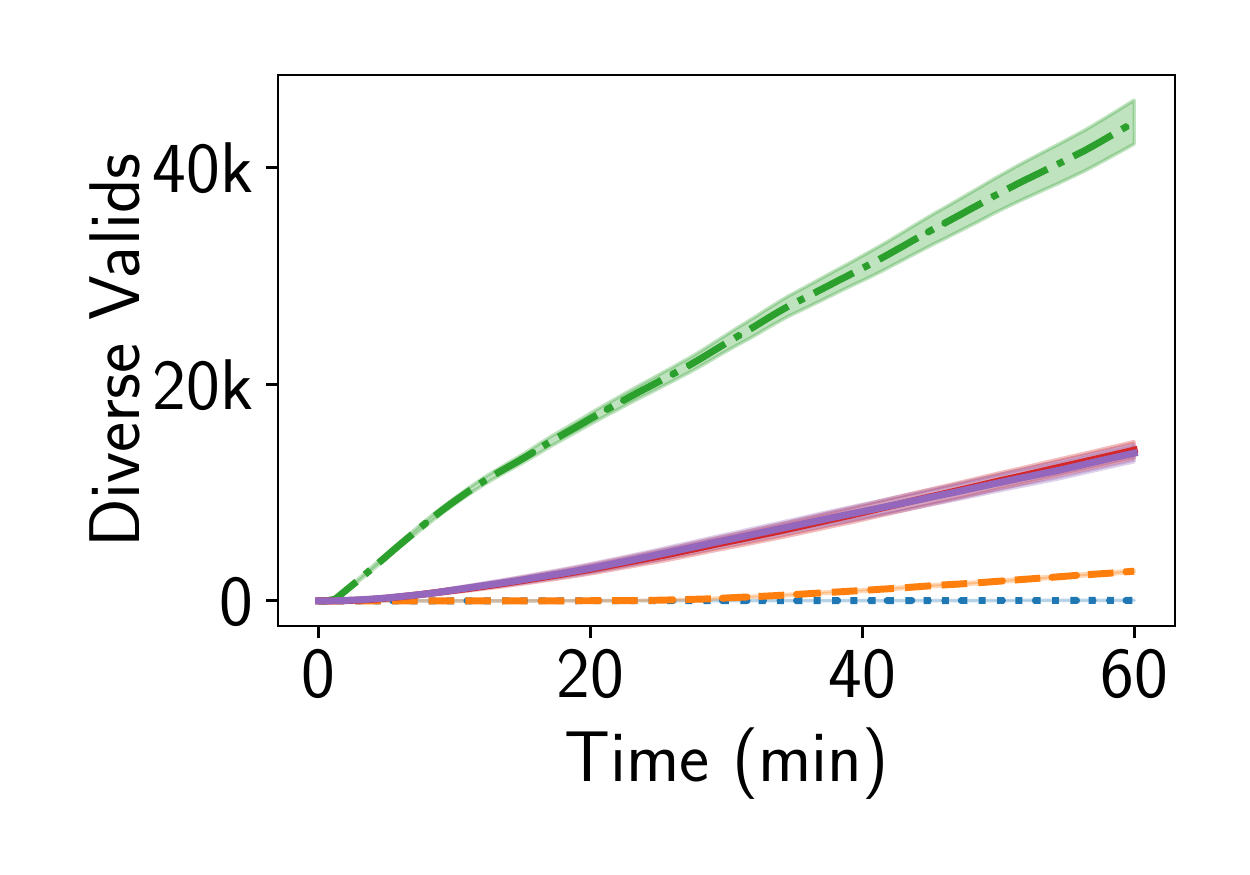}
	} \\[-1ex]
	\captionsetup[subfloat]{captionskip=-6pt}
	\subfloat[Maven]{
		\includegraphics[width = \divFigureWidth]{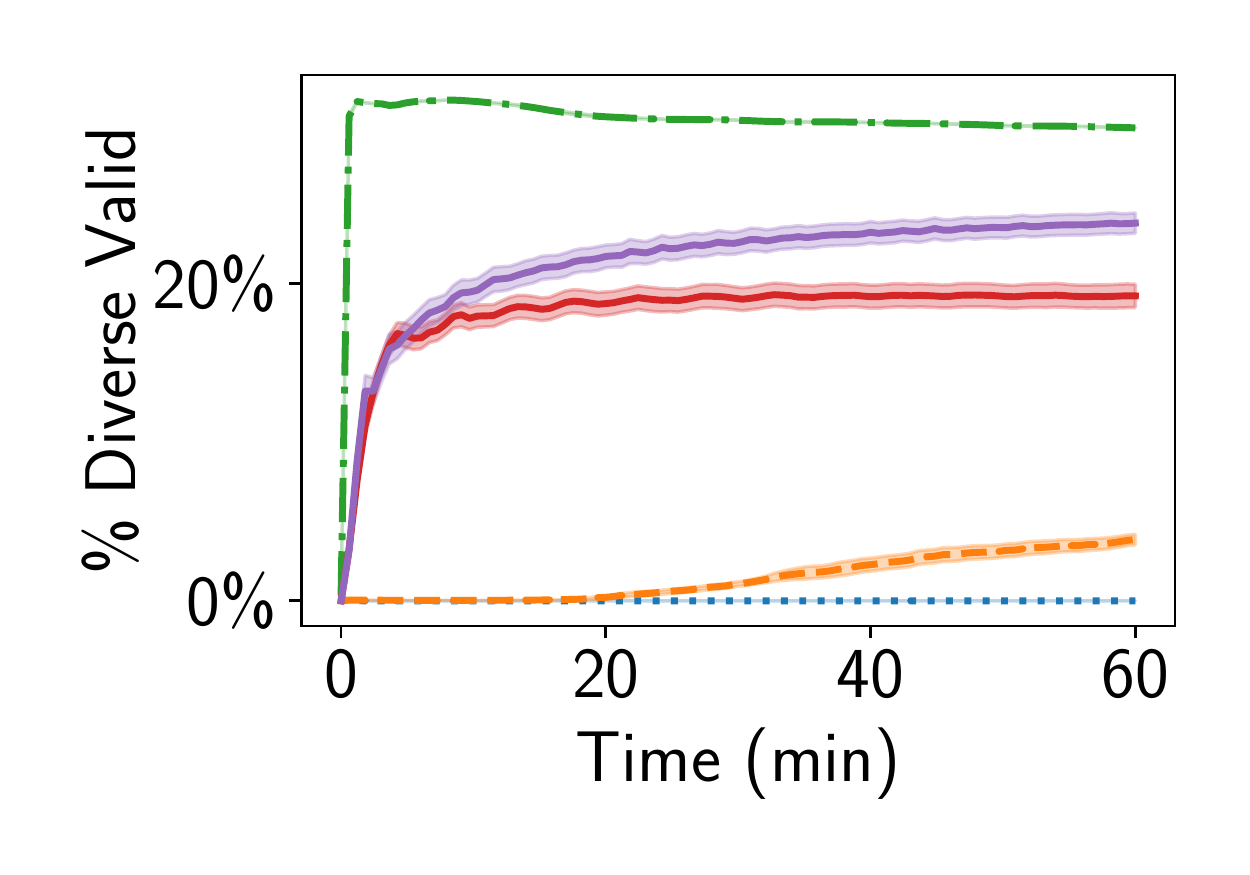}
		\includegraphics[width = \divFigureWidth]{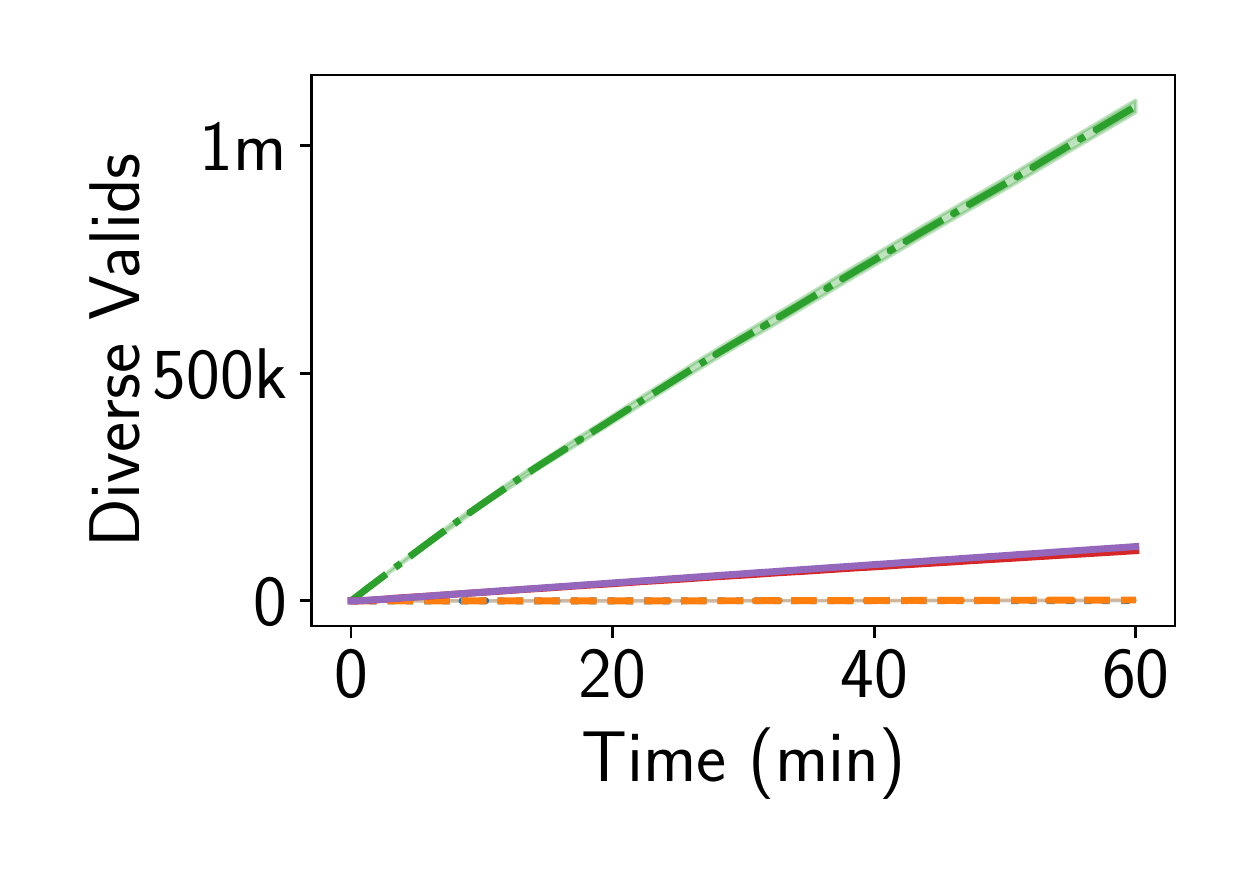}
	} \\[-1ex]
	\captionsetup[subfloat]{captionskip=-6pt}
	\subfloat[Rhino]{
		\includegraphics[width = \divFigureWidth]{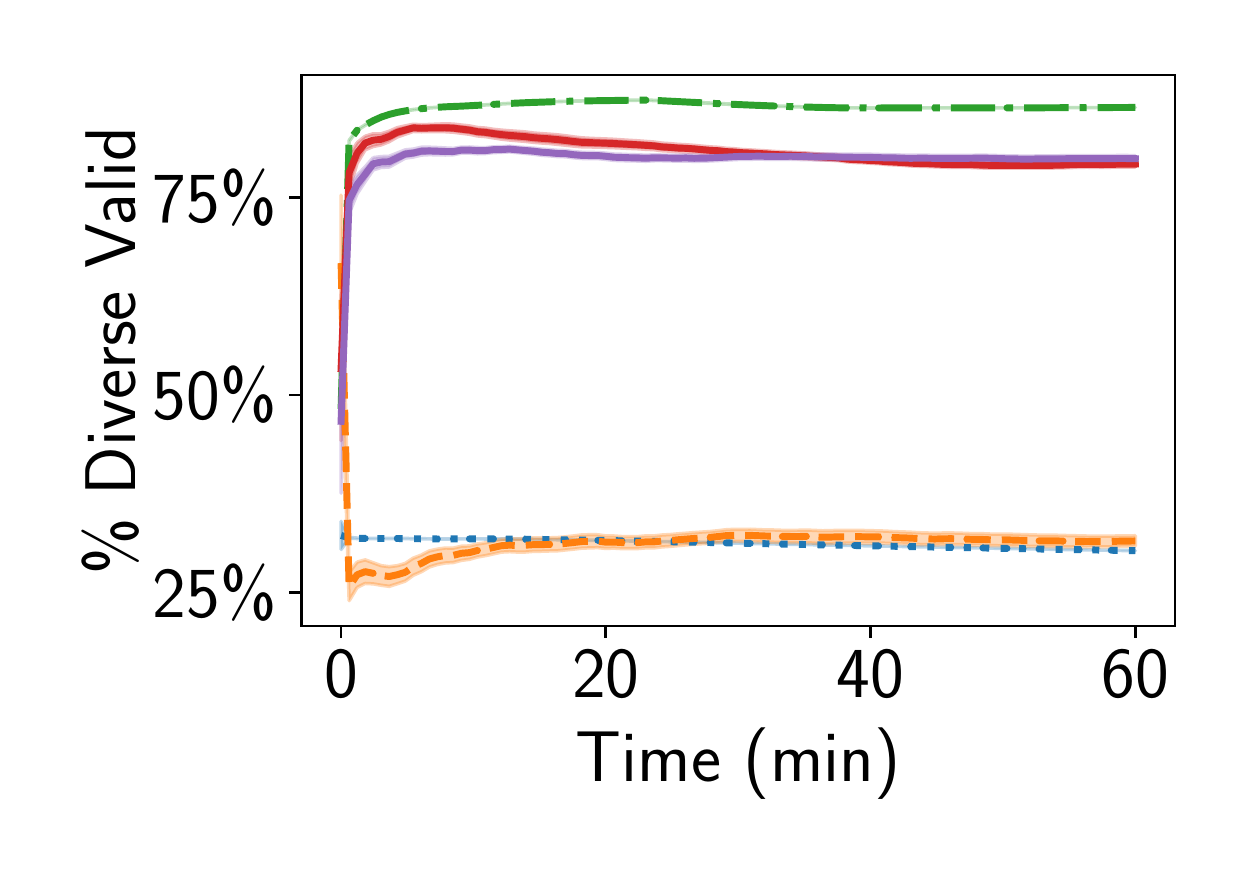}
		\includegraphics[width = \divFigureWidth]{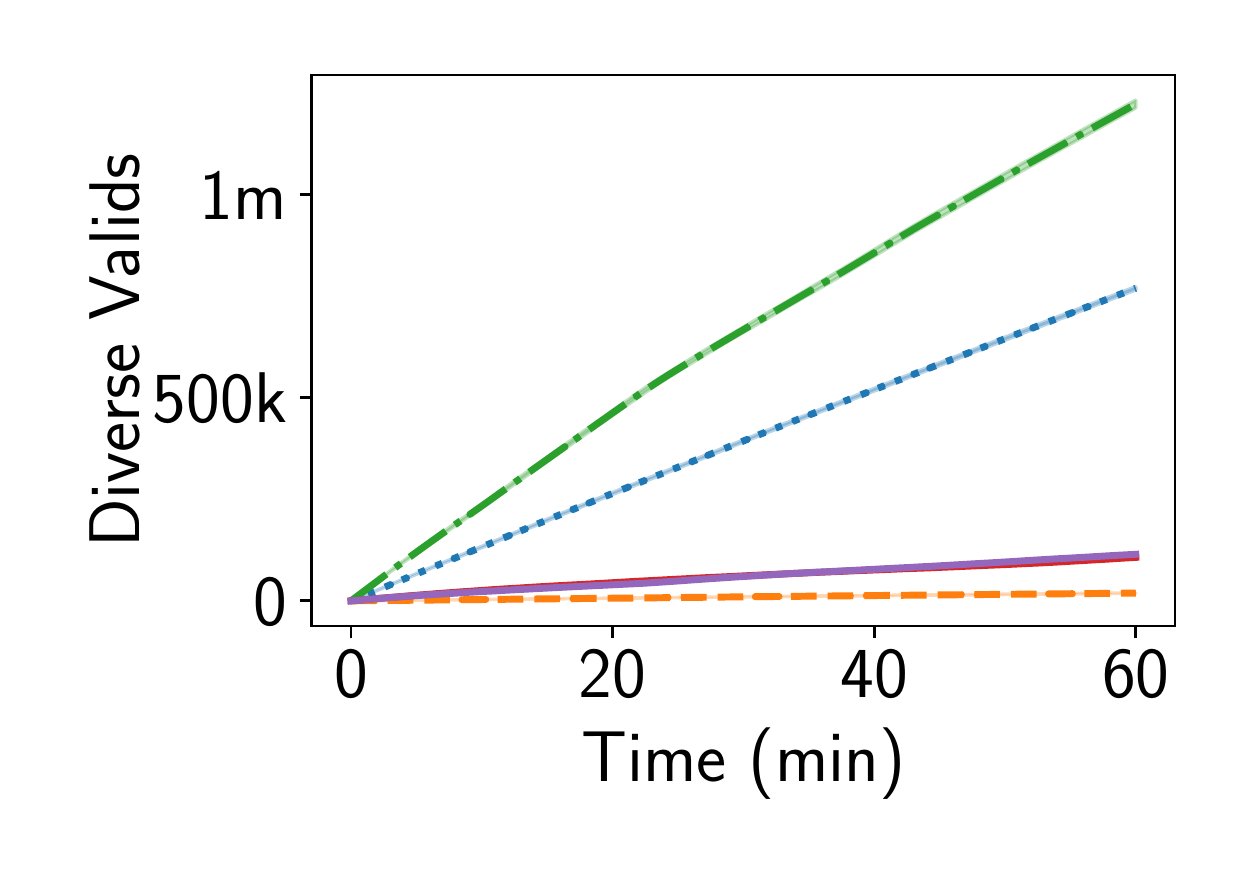}
	} \\[-1ex]
	\captionsetup[subfloat]{captionskip=-6pt}
	\subfloat[Closure]{
		\includegraphics[width = \divFigureWidth]{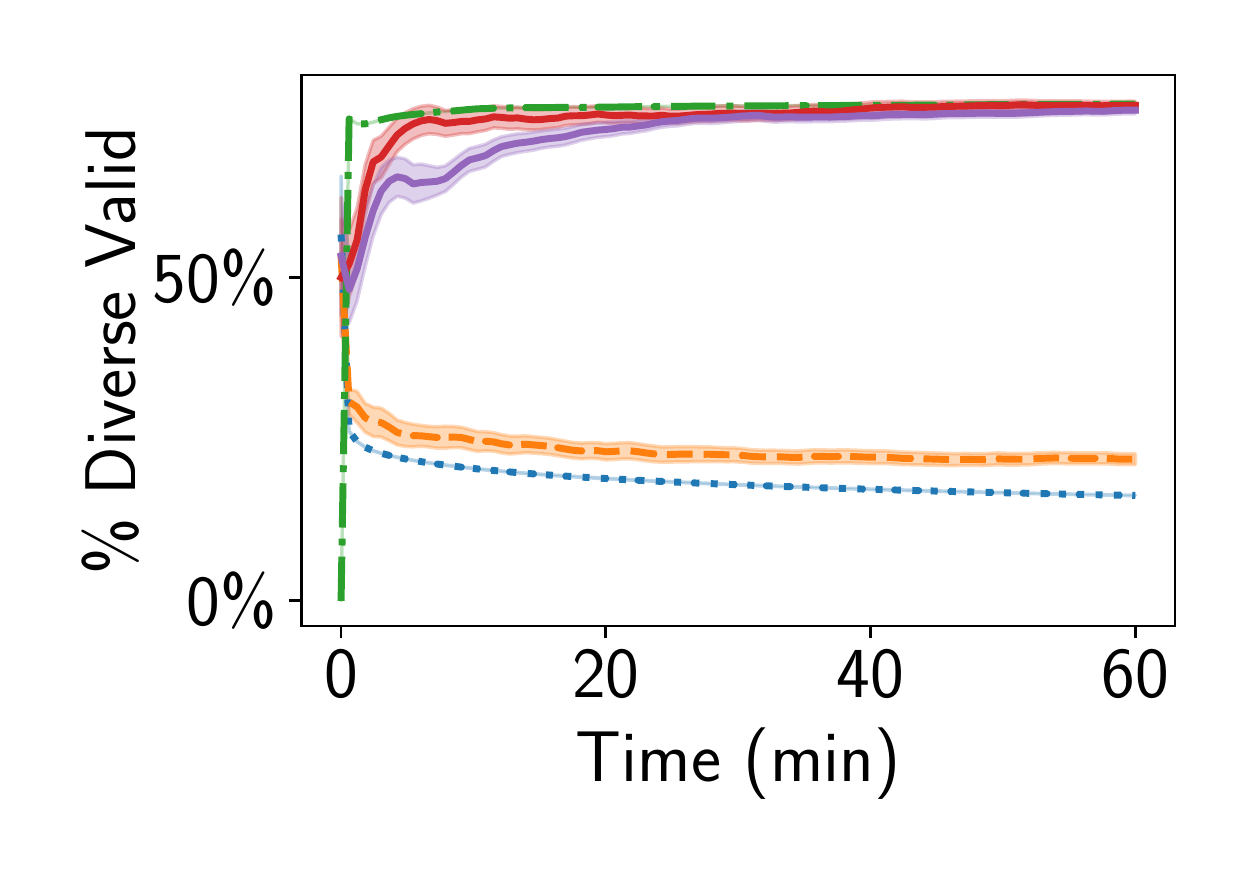}
		\includegraphics[width = \divFigureWidth]{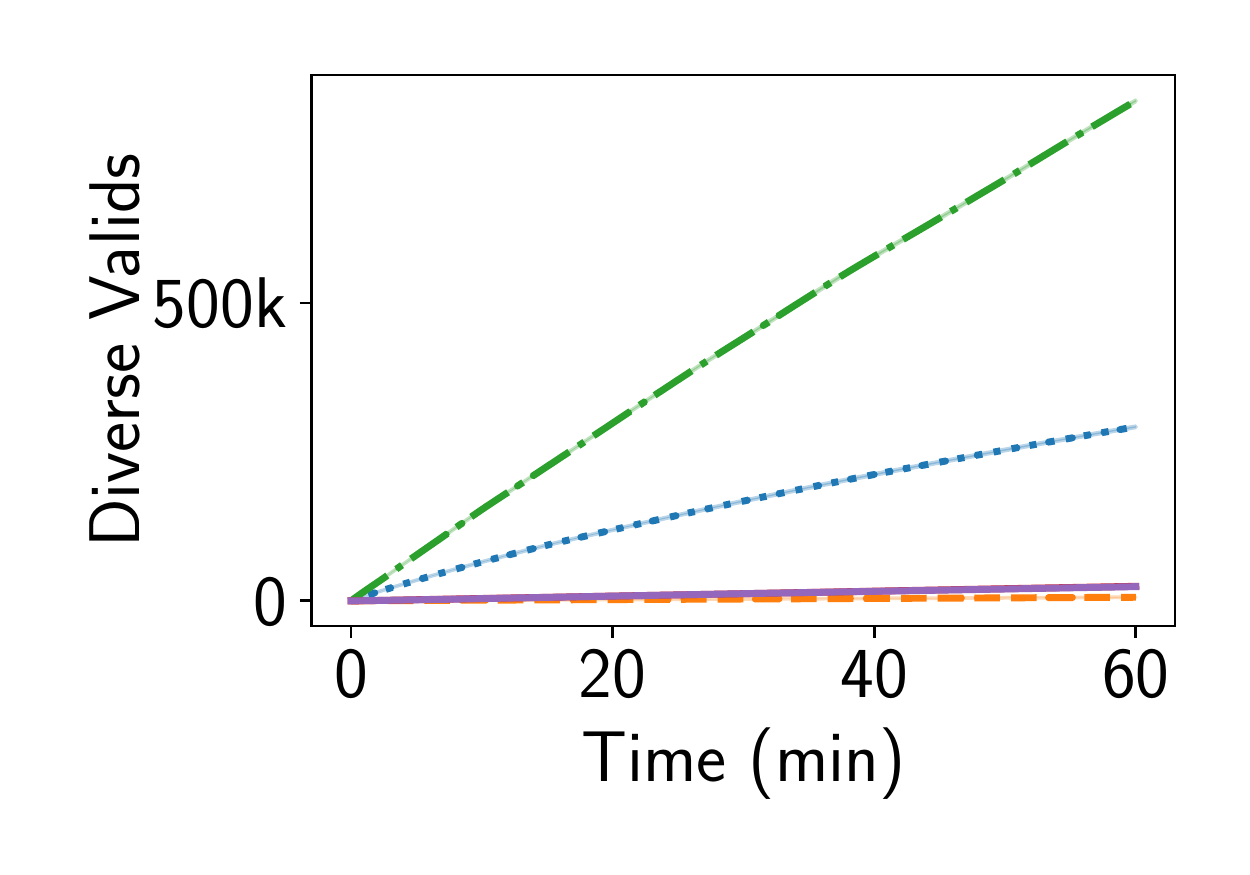}
	} \\[-1ex]
	\captionsetup[subfloat]{captionskip=-6pt}
	\subfloat[Nashorn]{
		\includegraphics[width = \divFigureWidth]{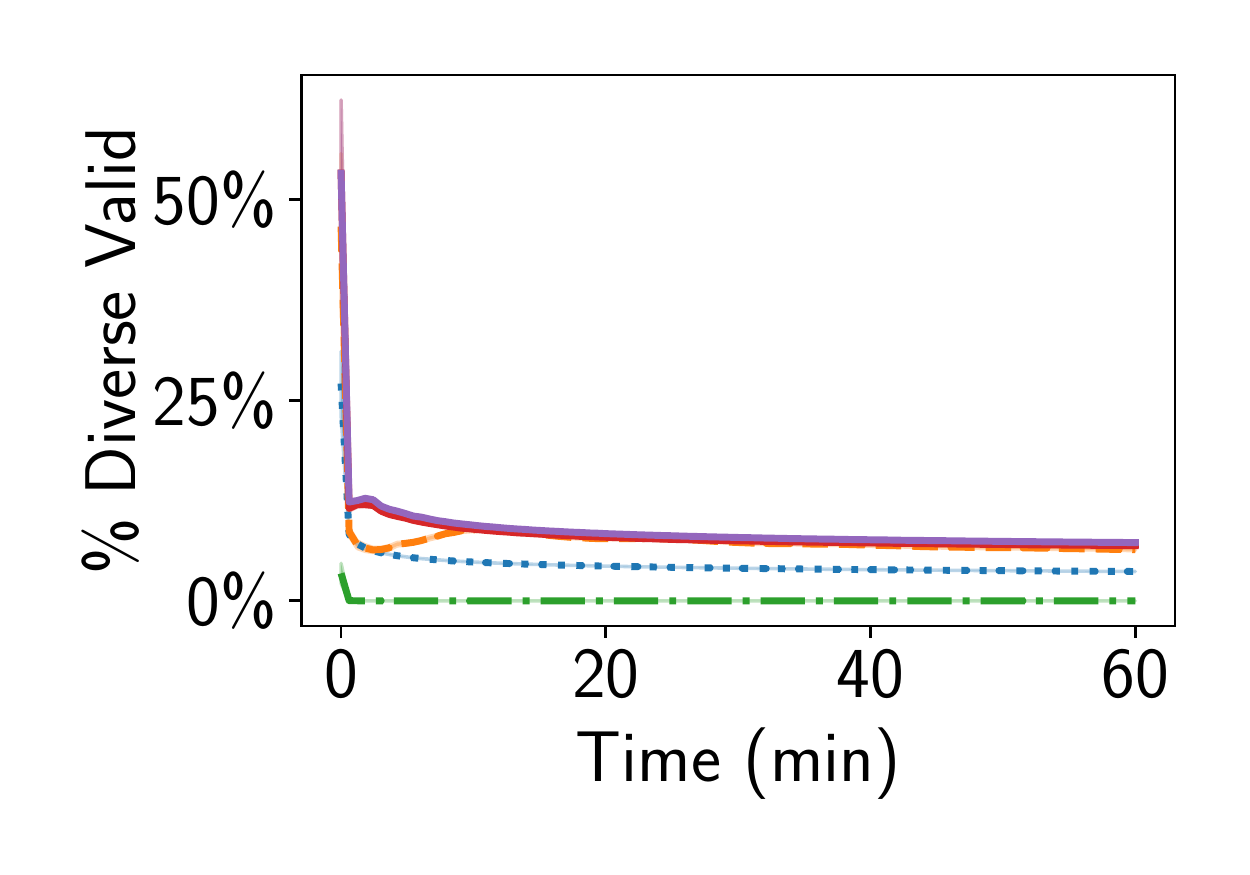}
		\includegraphics[width = \divFigureWidth]{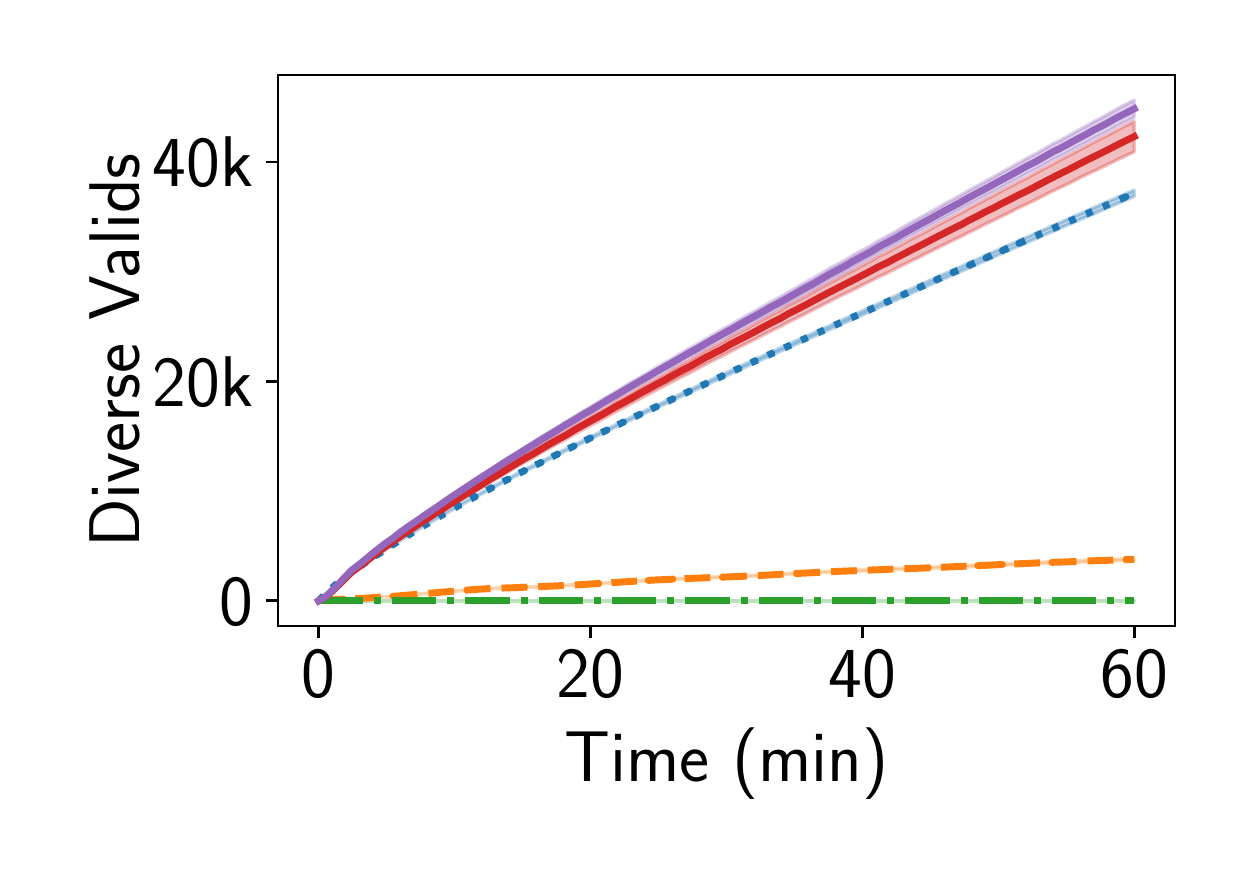}
	} \\[-1ex]
	\captionsetup[subfloat]{captionskip=-6pt}
	\subfloat[Tomcat]{
		\includegraphics[width = \divFigureWidth]{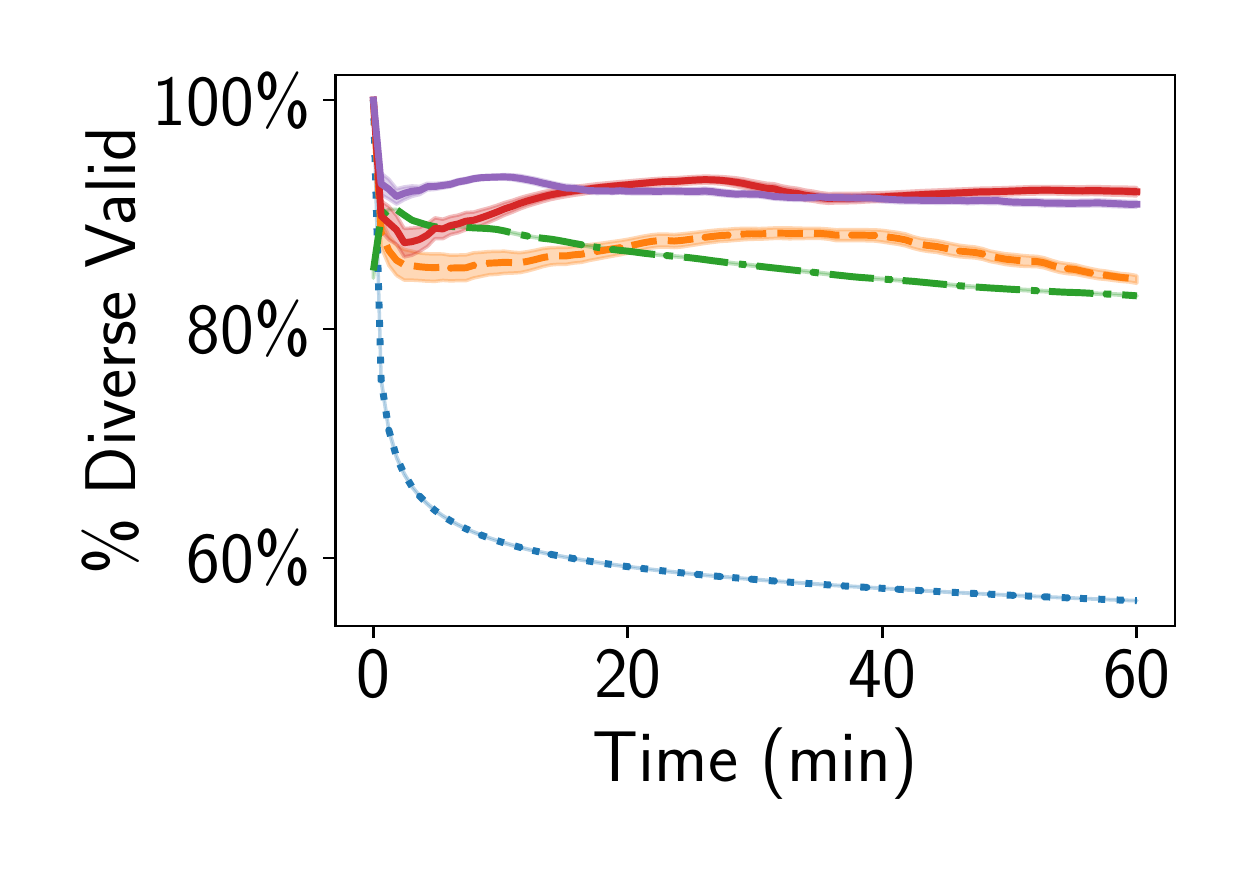}
		\includegraphics[width = \divFigureWidth]{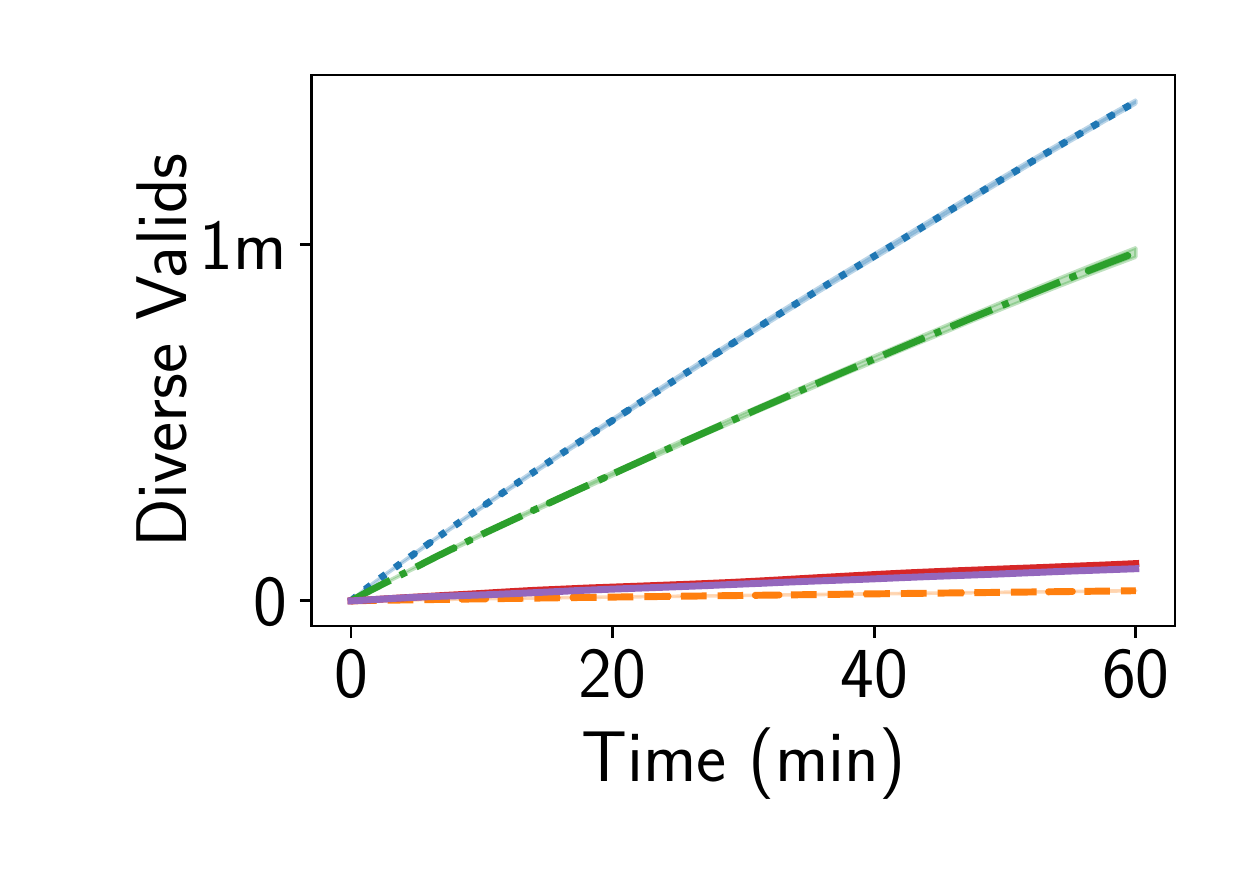}
	}
\end{tabular}
\caption{Percent (left) and absolute number (right) of diverse valid inputs (i.e., valid inputs with different traces).}
\label{fig:evaluation_divs}
\end{figure}

We answer \hyperref[sec:evaluation:RQ1]{RQ1} by assessing
the number of diverse valid inputs generated by \textsc{\ToolName}
compared to the baseline techniques.
Similar to Reddy et al. ~\cite{reddy2020rlcheck}, we consider \textsl{diverse valid inputs} as
inputs that exercise different traces, rather than inputs that
are only unique on the byte or string level.
The results are visualized in Figure \ref{fig:evaluation_divs}.
The left column shows for each benchmark subject
the \textsl{percentage} of all generated inputs
that were diverse valid inputs, whereas
the right column depicts the \textsl{total} number
of diverse valid inputs.

If we consider the percentages of generated diverse valid inputs,
the results indicate that \textsc{\ToolName} is competitive
with the current state-of-the-art.
For all benchmark subjects, both configurations of \textsc{\ToolName} outperform Zest and Quickcheck
as they are able to maintain a high percentage of diverse valid inputs and
even increase it over time for several subjects.
However, in two out of six subjects,
RLCheck significantly outperforms \textsc{\ToolName},
which is most notable in the Maven benchmark
where the mean percentage of diverse valid inputs is about $6\%$ and $10\%$
higher compared to \textsc{\ToolName-structure} and \textsc{\ToolName-simple}, respectively.

The plots for Ant show a high variability in the performance of \textsc{\ToolName},
which is due to the difficulty of finding the first valid input for this subject.
This indicates that \textsc{\ToolName} can potentially benefit from an initial valid seed input.
The same applies to Zest, which also further mutates invalid inputs by default, thus 
potentially wasting resources by initially exploring the error handling code of the SUT.
On the other hand, the efficiency of RLCheck allows it to quickly find a valid input and exploit the obtained information
to generate further diverse valid inputs.

On our newly added subjects Nashorn and Tomcat, RLCheck does not perform as well as on the other benchmarks.
This can potentially be attributed to the fact that
the new subjects have very simple validity functions, while different behavior may only be triggered
by specific input structures.
We assume that since RLCheck only relies on the validity feedback, it is unable to learn any meaningful policy
as almost any generated input is valid, but not necessarily diverse valid.
As a result, RLCheck might be prone to overfit to a valid, yet
uninteresting space of inputs (with regard to the triggered program behavior).
In contrast to that, \textsc{\ToolName} also utilizes code coverage to
identify whether an input with a unique structure exercises interesting behavior or not.

While the results indicate that \textsc{\ToolName} is highly effective in generating diverse valid inputs,
the total numbers (Figure \ref{fig:evaluation_divs}, right column) show
that our approach is mostly limited by its efficiency.
Overall, \textsc{\ToolName} only performs better than Zest in terms of
generating many diverse valid inputs.
When comparing both configurations of \textsc{\ToolName},
\textsc{\ToolName-structure} tends to have a slight edge
due to the added analysis of input structures, but
the differences are generally not significant.
However \textsc{\ToolName}, is outperformed by both RLCheck and Quickcheck
for most of the benchmark subjects.
This can be explained as follows.
Like Zest, \textsc{\ToolName} requires the SUT to
be instrumented in order to collect code coverage.
This results in a significant slowdown of execution time (around $10$--$100\times$ slower),
effectively reducing the total number of diverse valid inputs that can potentially be generated.
On the other hand, while RLCheck has a comparable effectiveness to \textsc{\ToolName},
the blackbox approach makes it much more efficient.
Nevertheless, this is an acceptable trade-off,
since our main focus is not on producing a large number of diverse valid inputs.
Instead, the overall goal of \textsc{\ToolName} is to exercise \textsl{diverse behavior},
which we evaluate in the following section.

\subsection{RQ2: Diverse Execution of Program Behavior} \label{sec:results-RQ2}

\newcommand{\coverageFigureWidth}{1.75in}
\begin{figure*}[tpb]
\begin{tabular}{c}
	\includegraphics[width=5in]{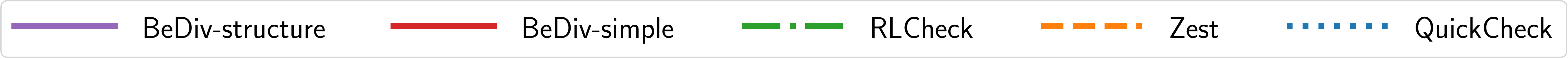} \\[-2ex]
	\captionsetup[subfloat]{captionskip=0pt}
	\subfloat[Ant]{
		\includegraphics[width = \coverageFigureWidth]{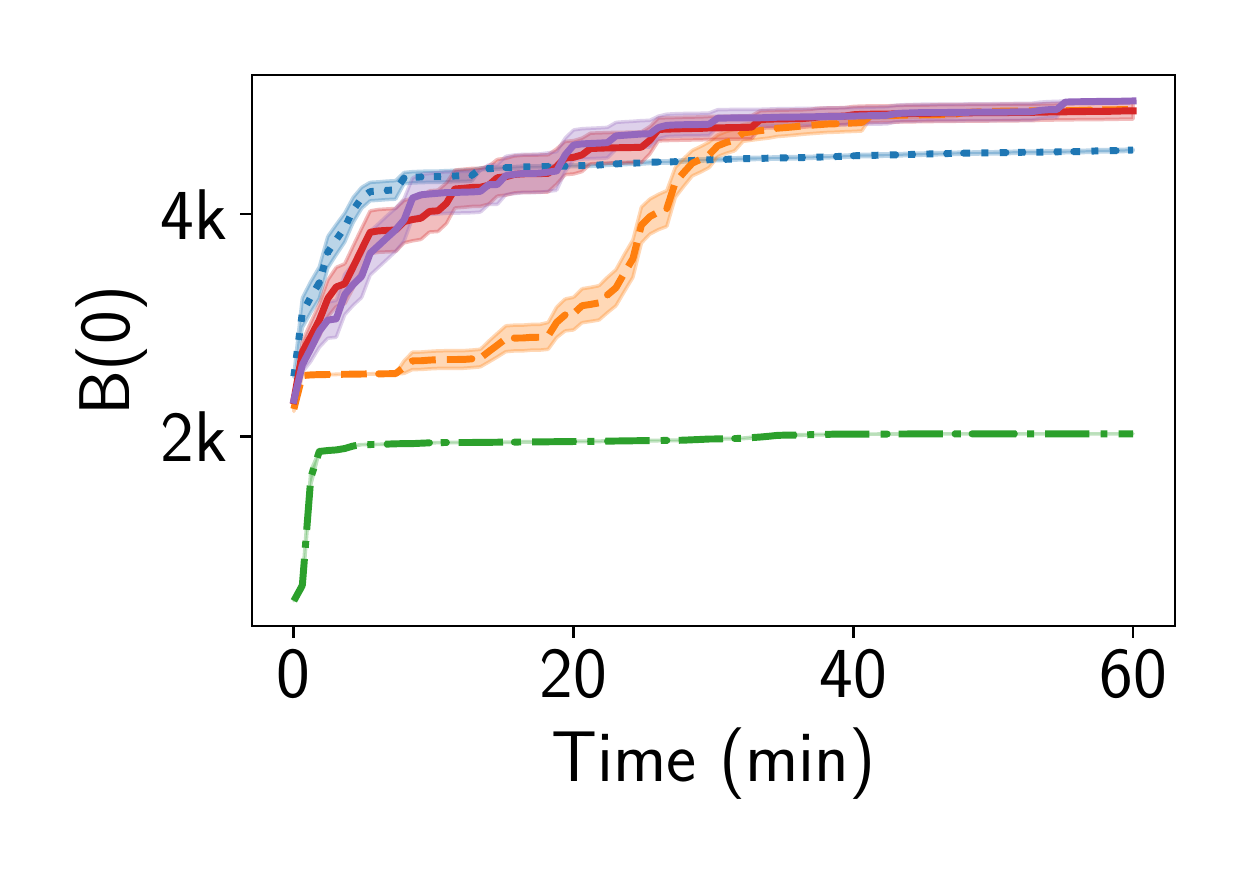}
		\includegraphics[width = \coverageFigureWidth]{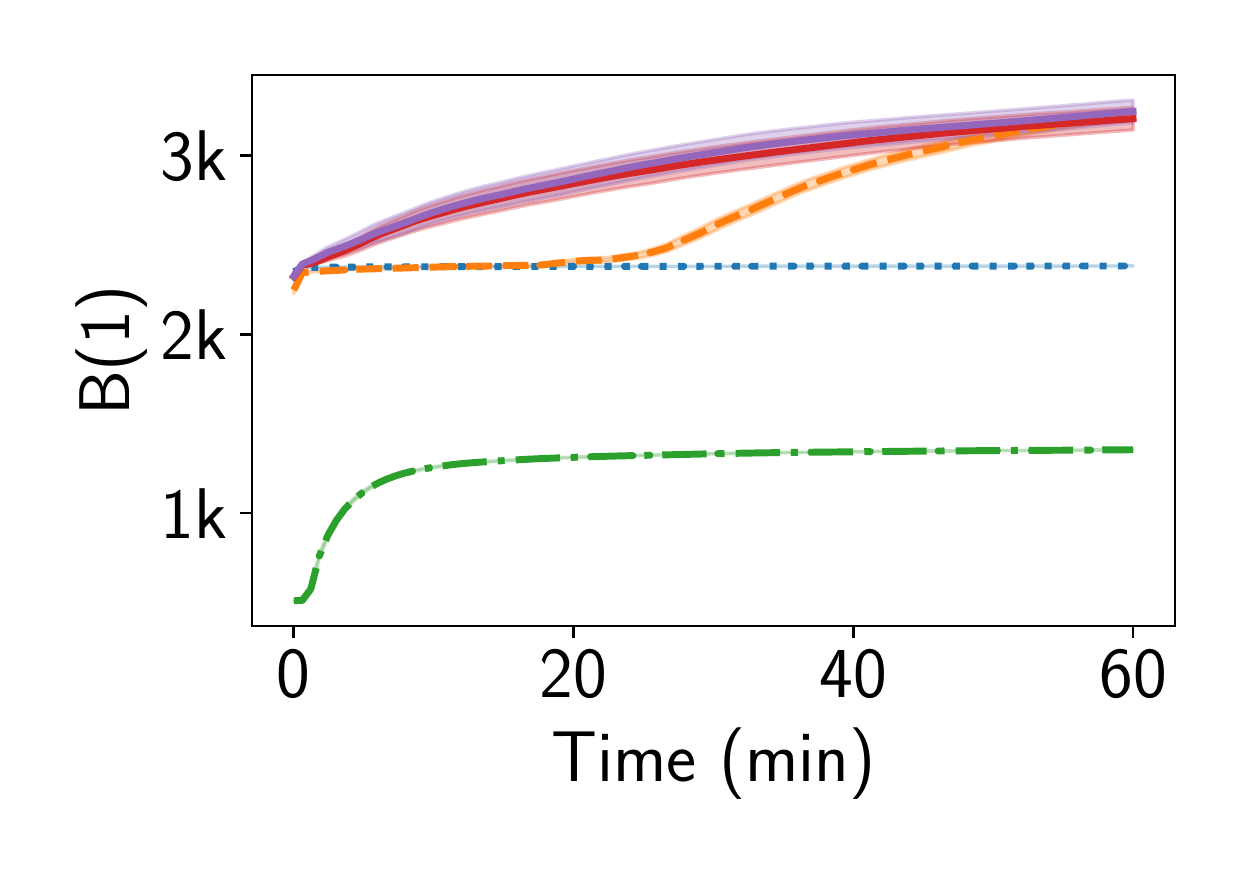}
		\includegraphics[width = \coverageFigureWidth]{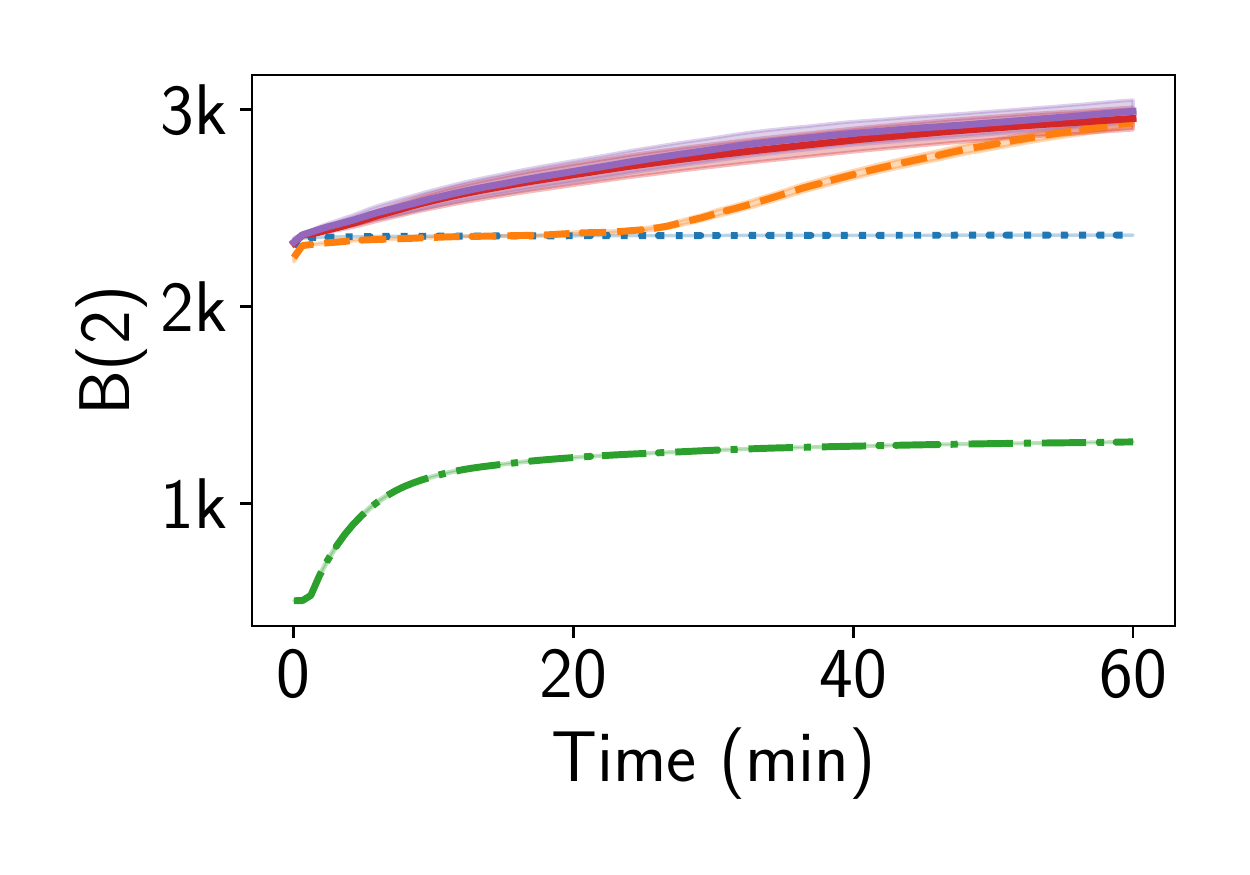}
	} \\[-2ex]
	\captionsetup[subfloat]{captionskip=0pt}
	\subfloat[Maven]{
		\includegraphics[width = \coverageFigureWidth]{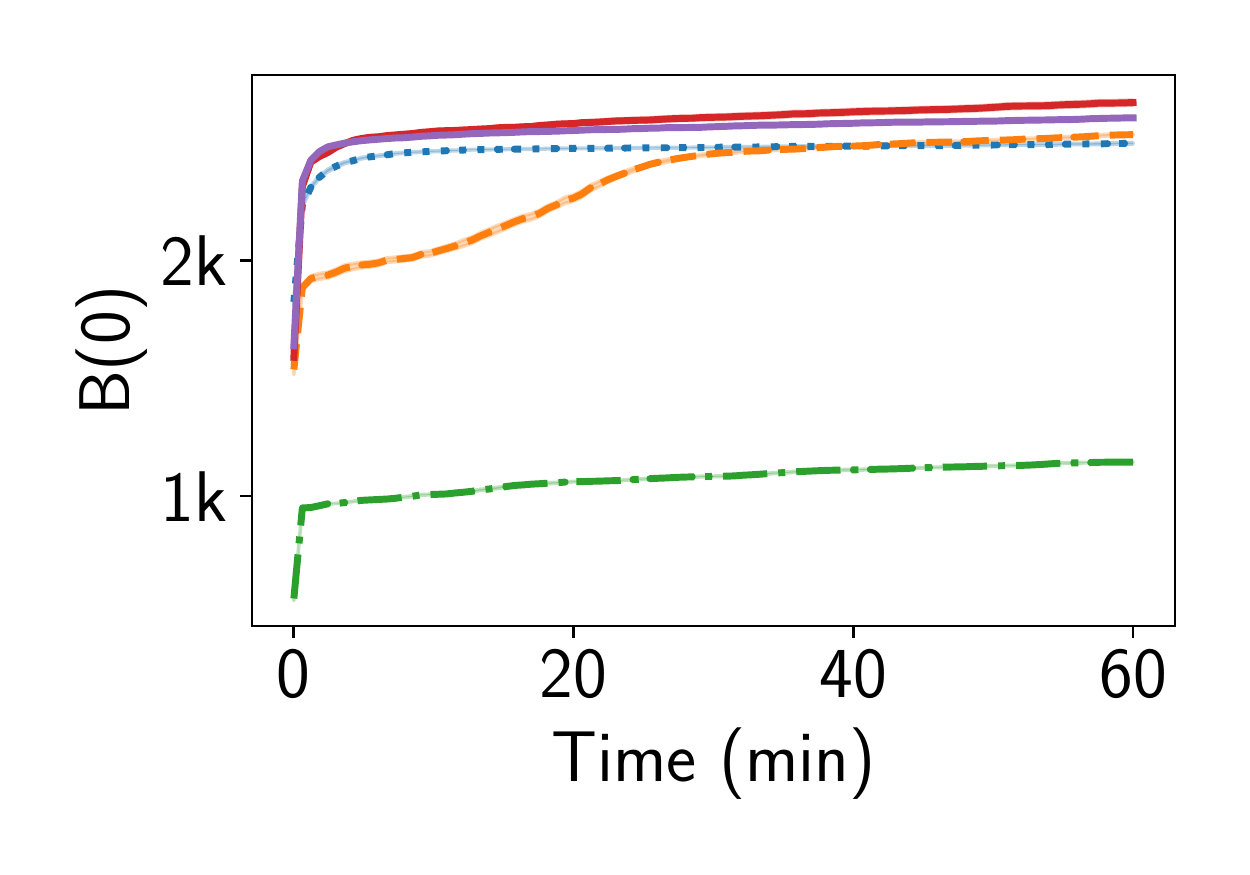}
		\includegraphics[width = \coverageFigureWidth]{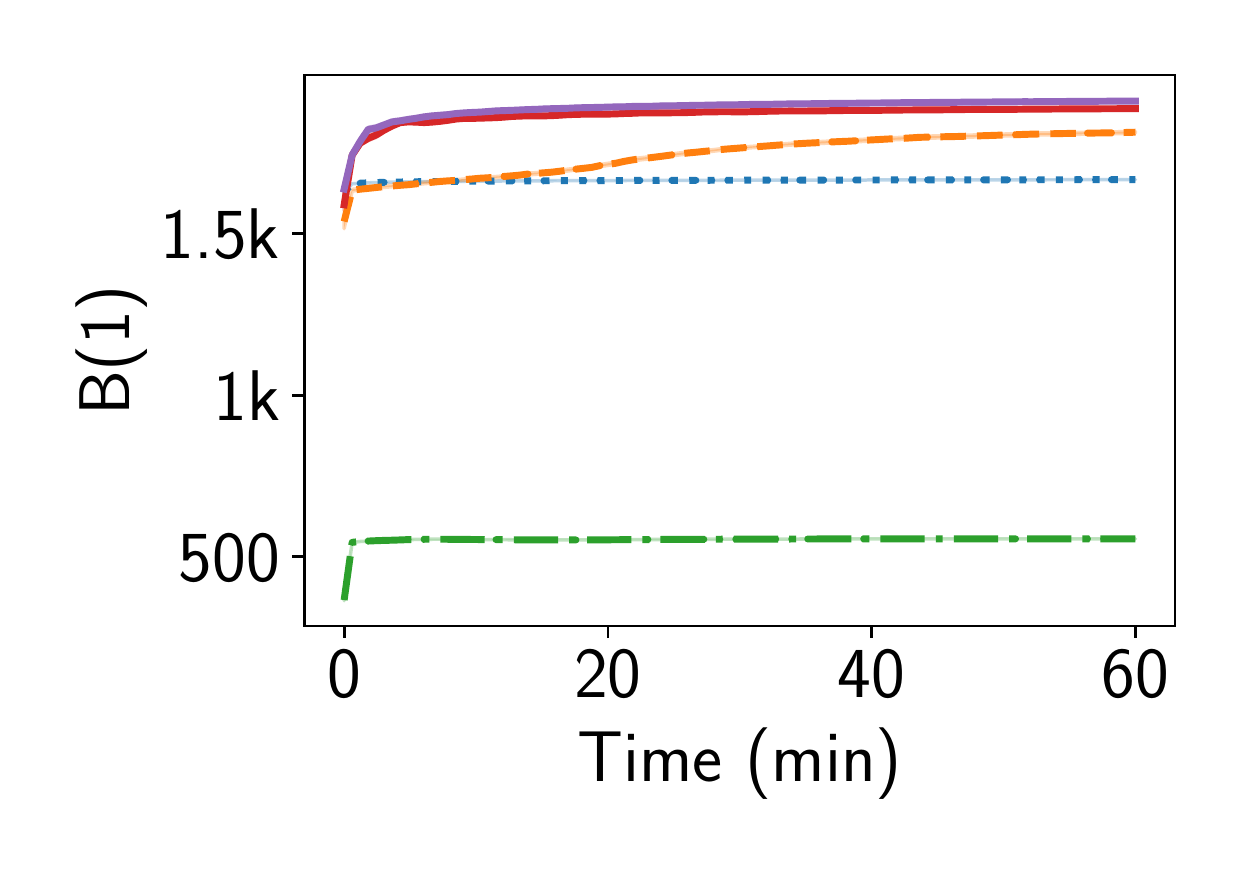}
		\includegraphics[width = \coverageFigureWidth]{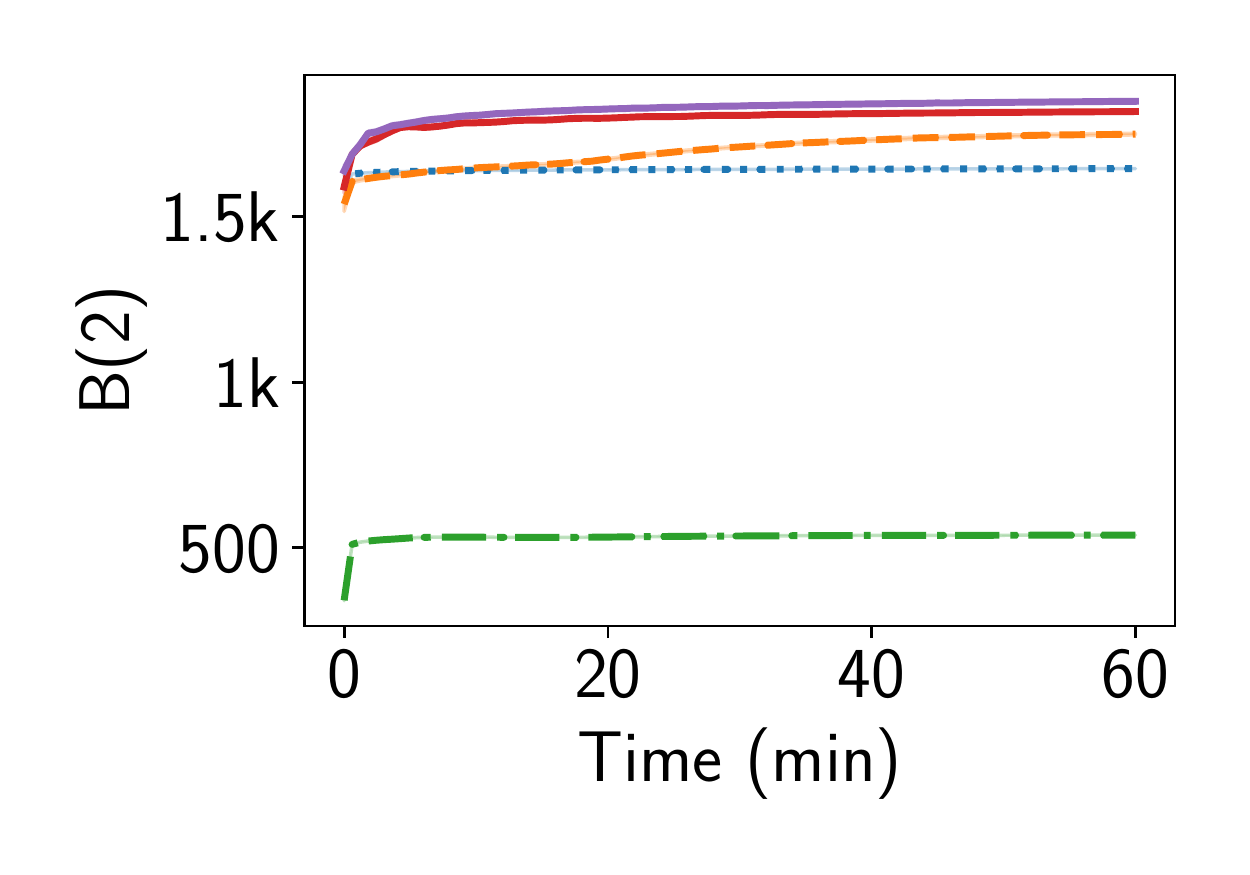}
	} \\[-2ex]
	\captionsetup[subfloat]{captionskip=0pt}
	\subfloat[Rhino]{
		\includegraphics[width = \coverageFigureWidth]{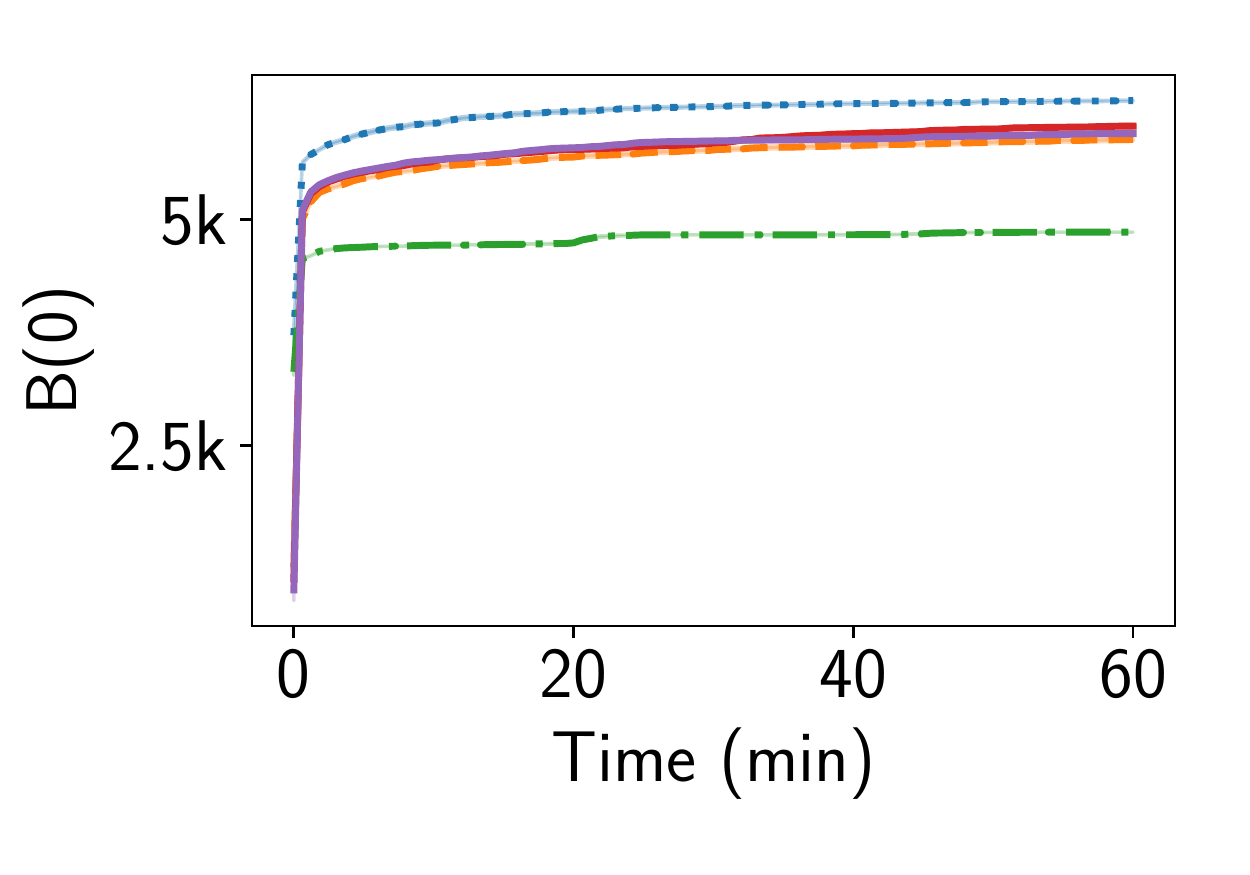}
		\includegraphics[width = \coverageFigureWidth]{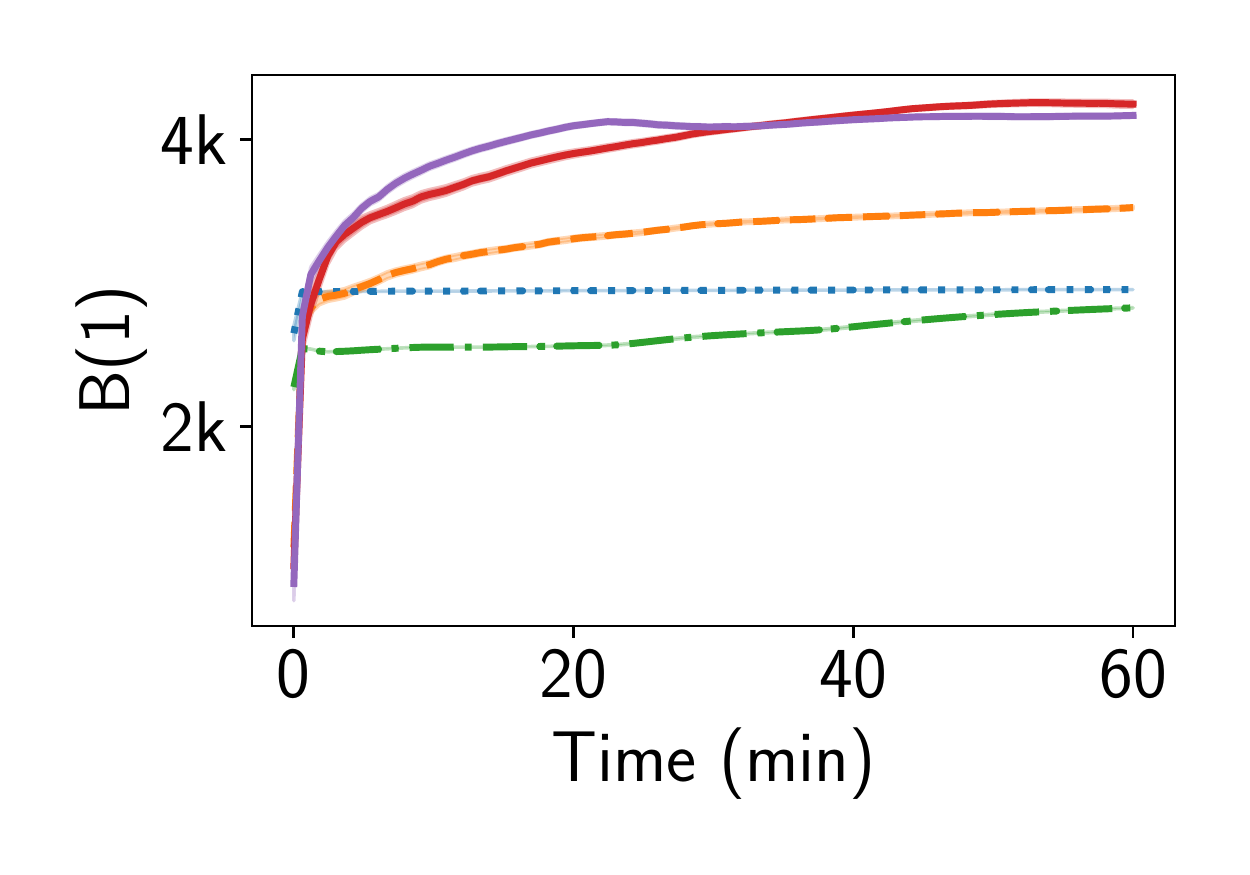}
		\includegraphics[width = \coverageFigureWidth]{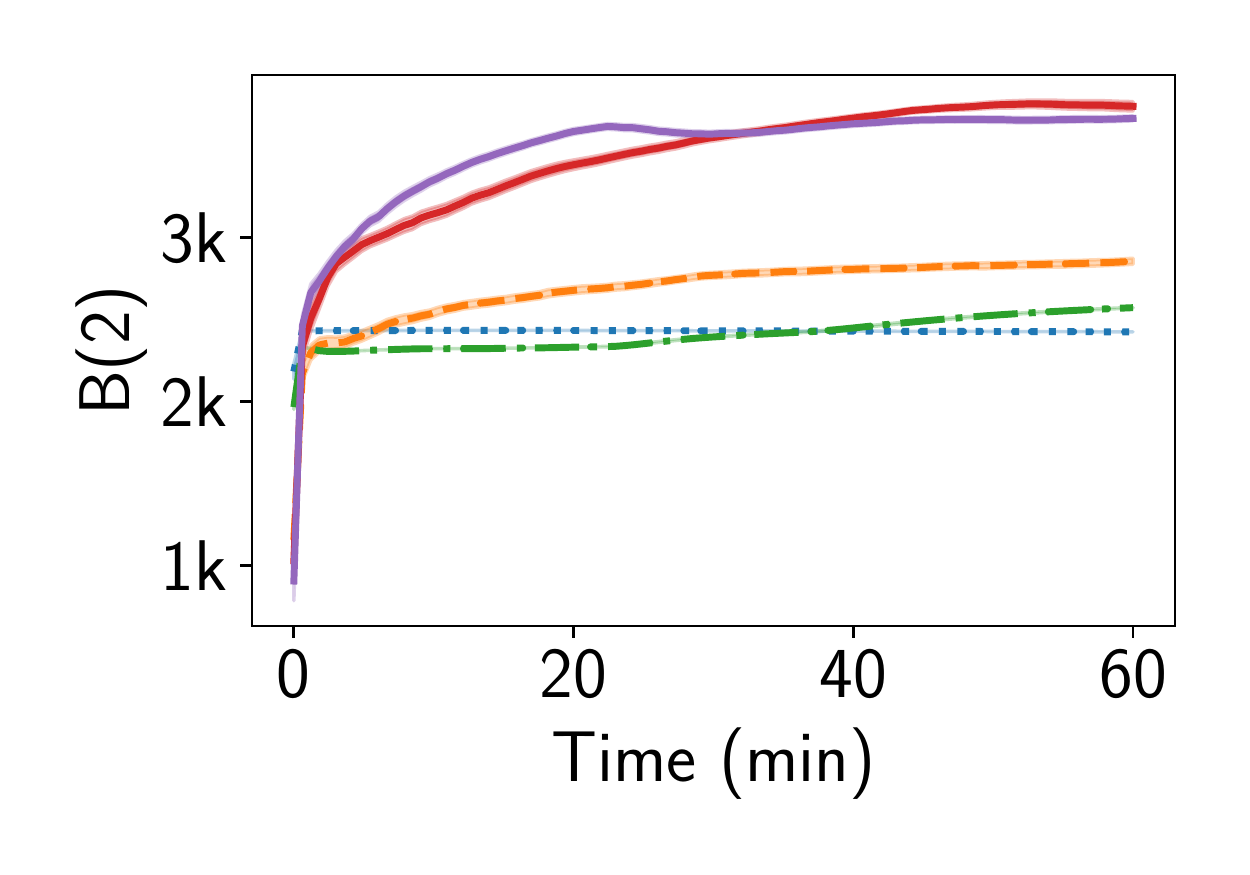}
	} \\[-2ex]
	\captionsetup[subfloat]{captionskip=0pt}
	\subfloat[Closure]{
		\includegraphics[width = \coverageFigureWidth]{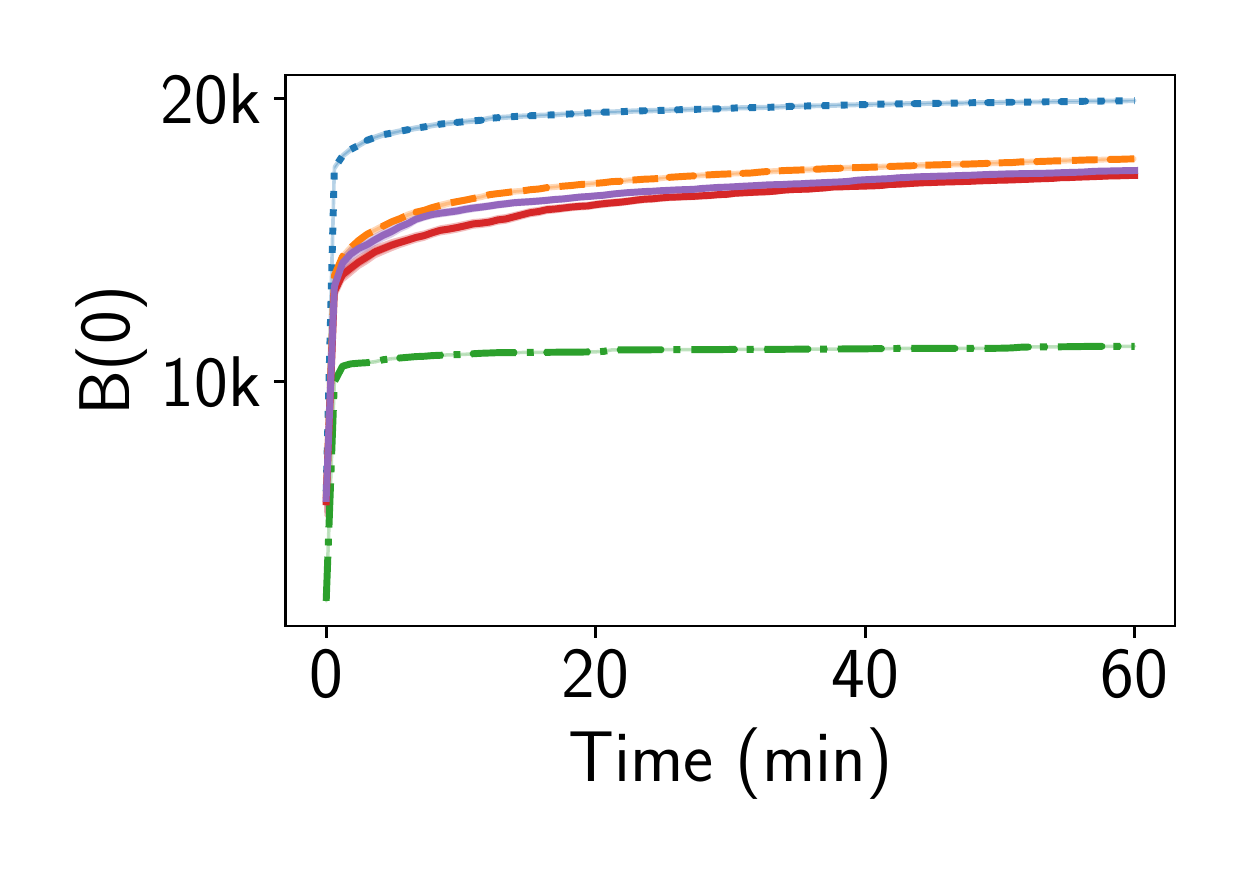}
		\includegraphics[width = \coverageFigureWidth]{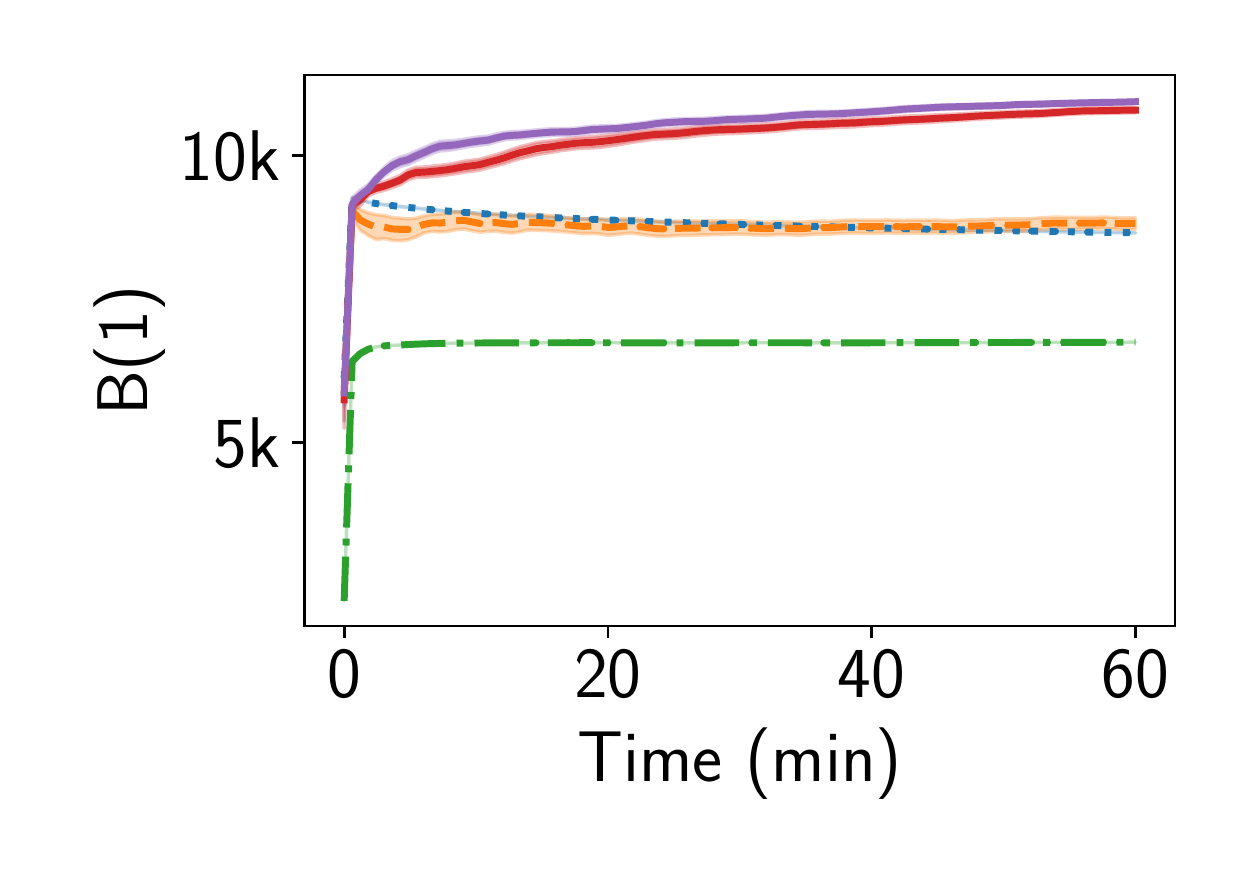}
		\includegraphics[width = \coverageFigureWidth]{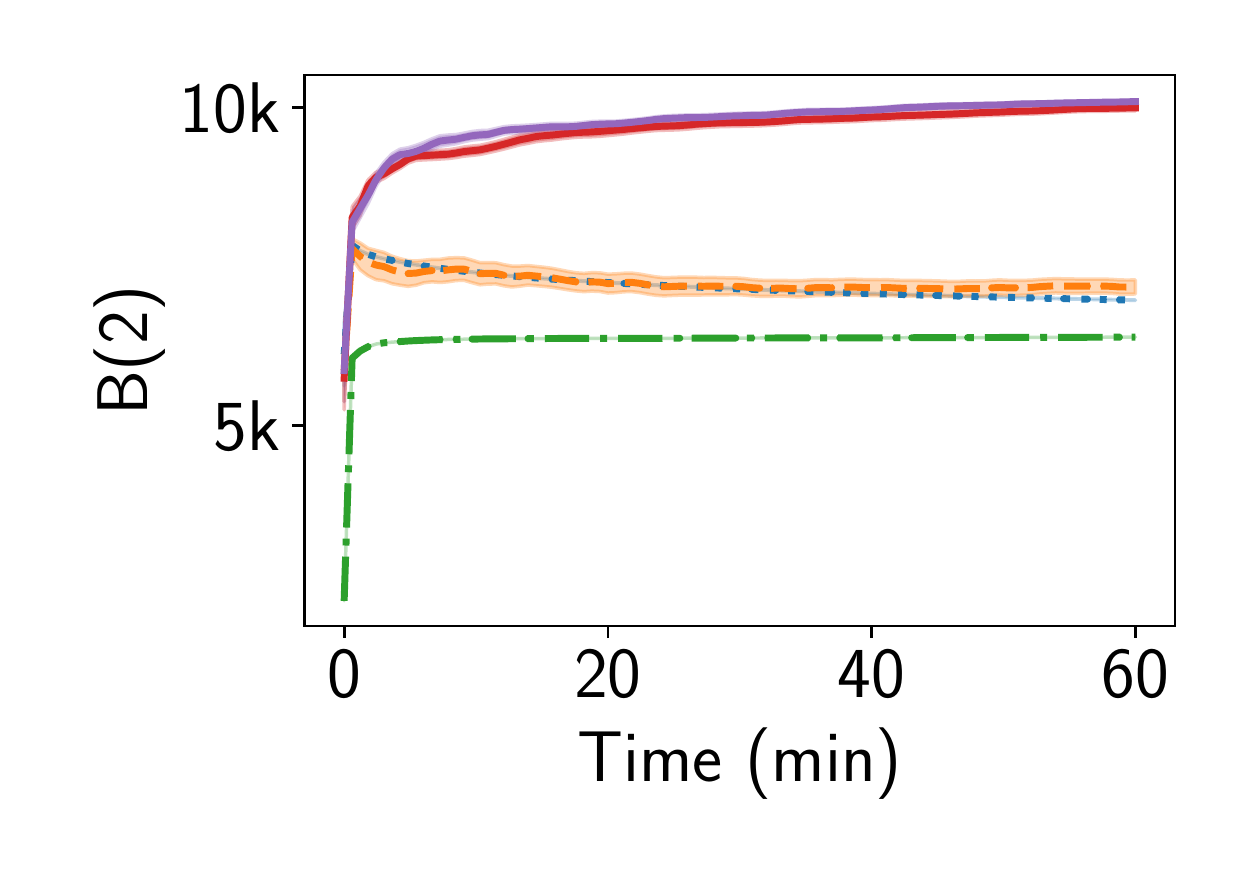}
	} \\[-2ex]
	\captionsetup[subfloat]{captionskip=0pt}
	\subfloat[Nashorn]{
		\includegraphics[width = \coverageFigureWidth]{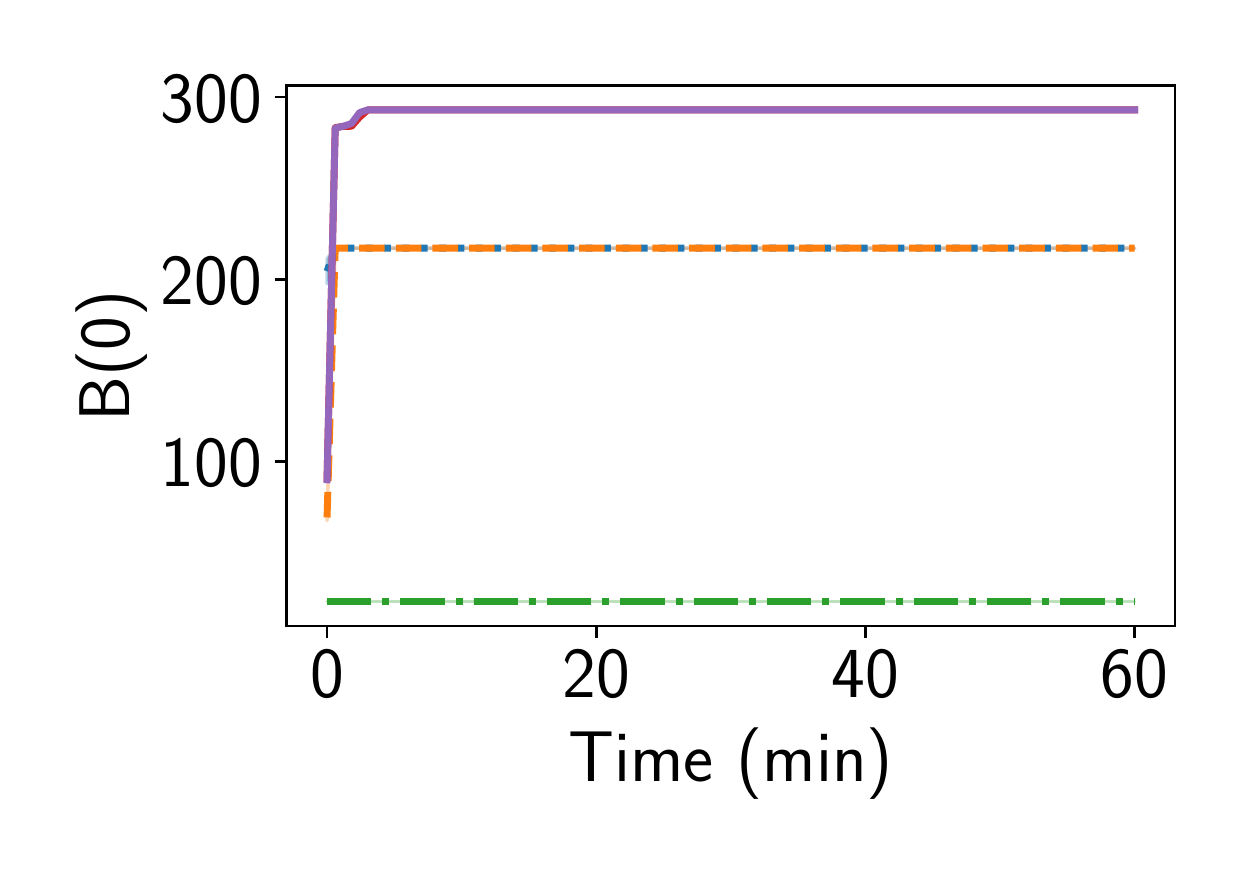}
		\includegraphics[width = \coverageFigureWidth]{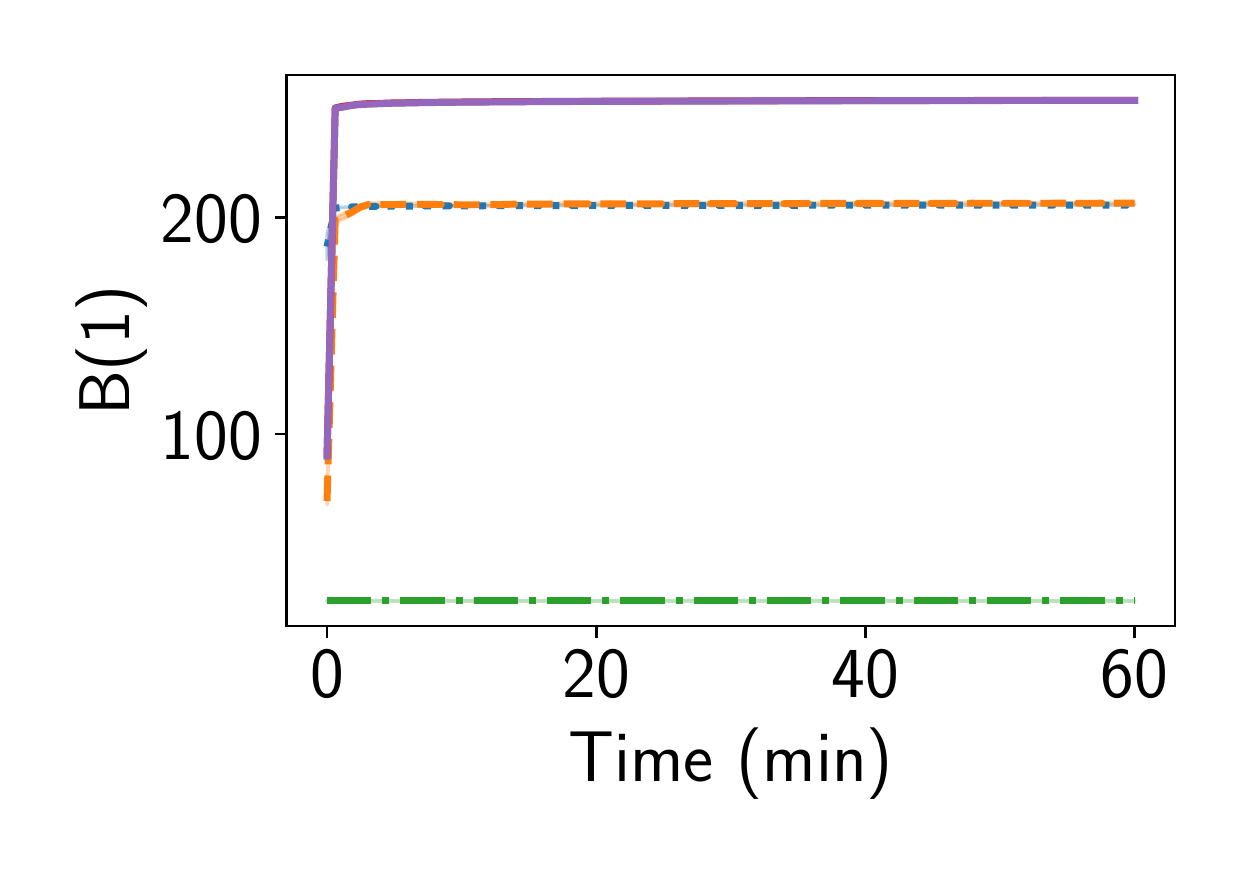}
		\includegraphics[width = \coverageFigureWidth]{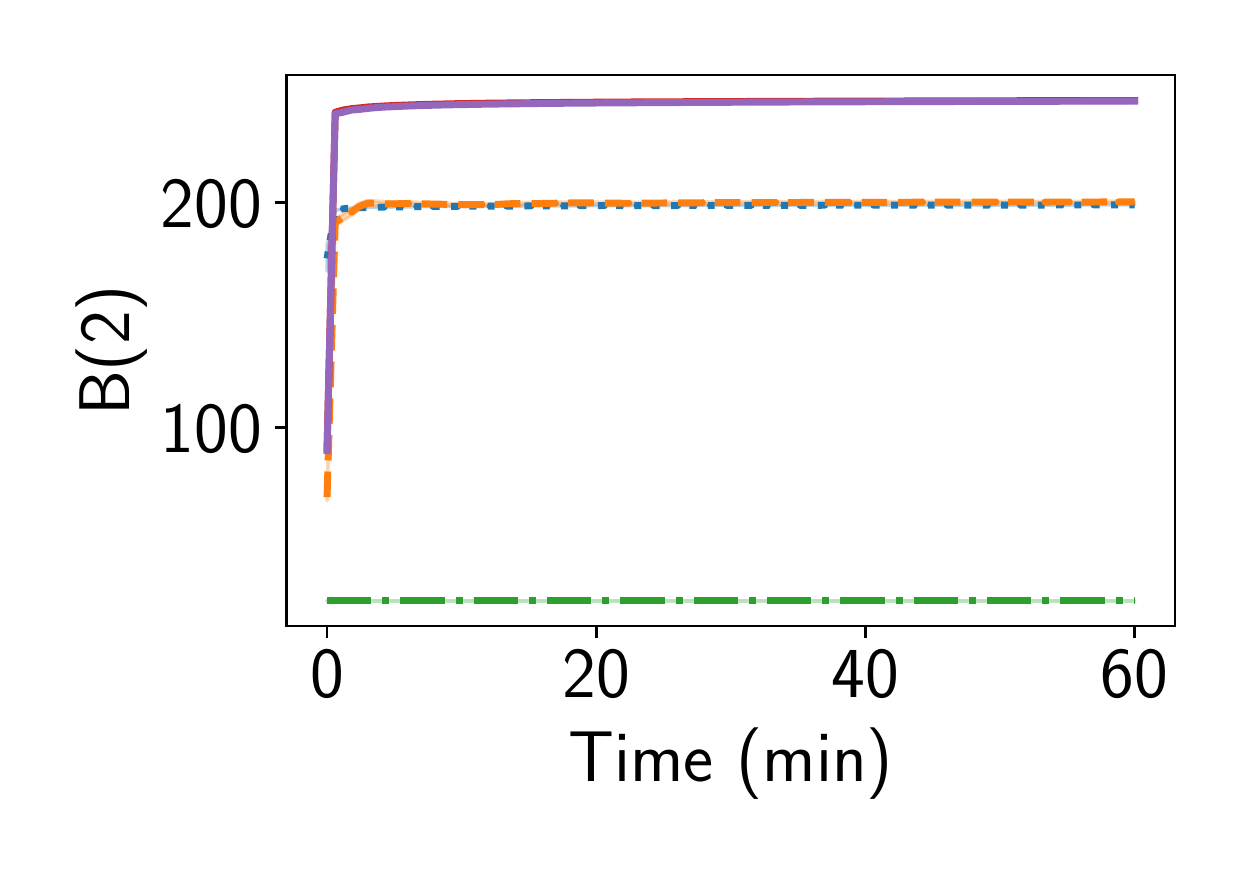}
	} \\[-2ex]
	\captionsetup[subfloat]{captionskip=0pt}
	\subfloat[Tomcat]{
		\includegraphics[width = \coverageFigureWidth]{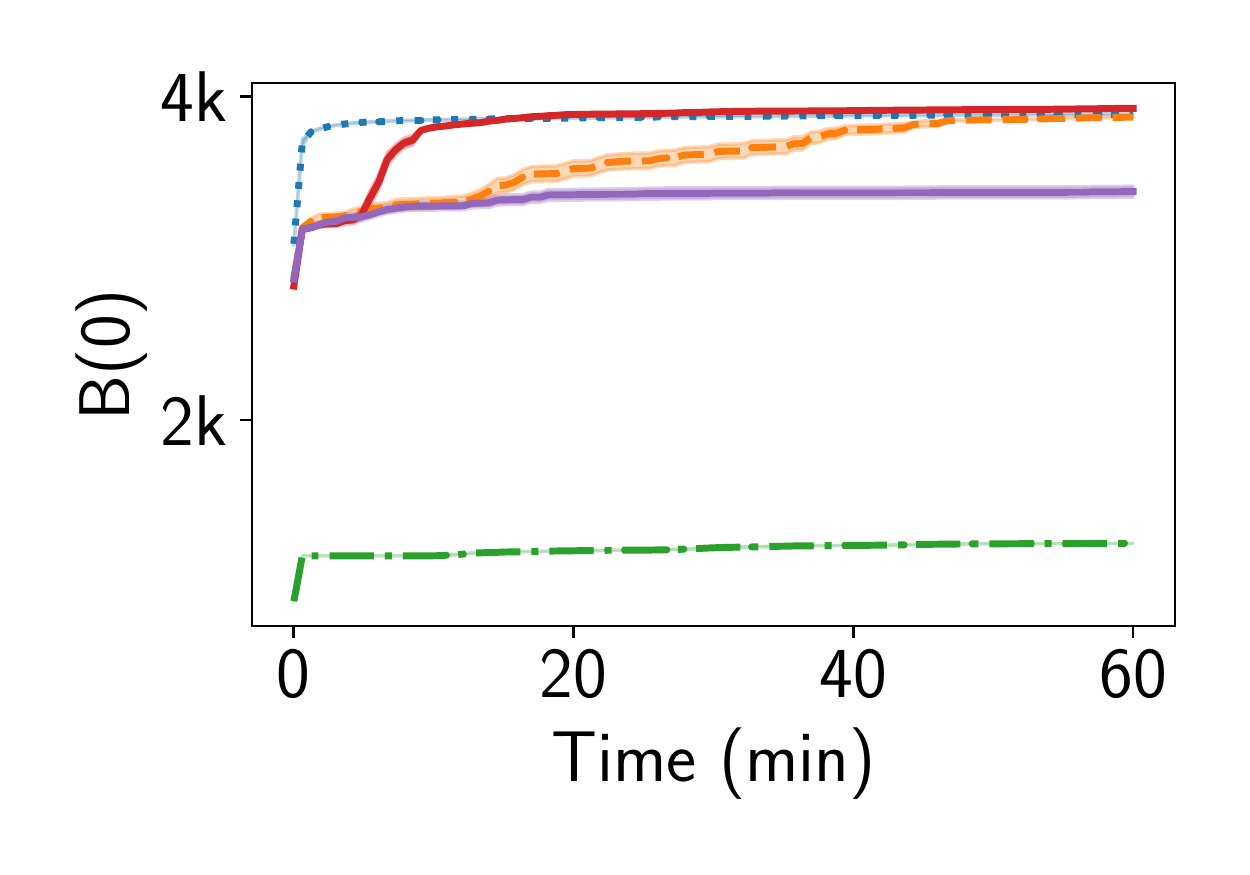}
		\includegraphics[width = \coverageFigureWidth]{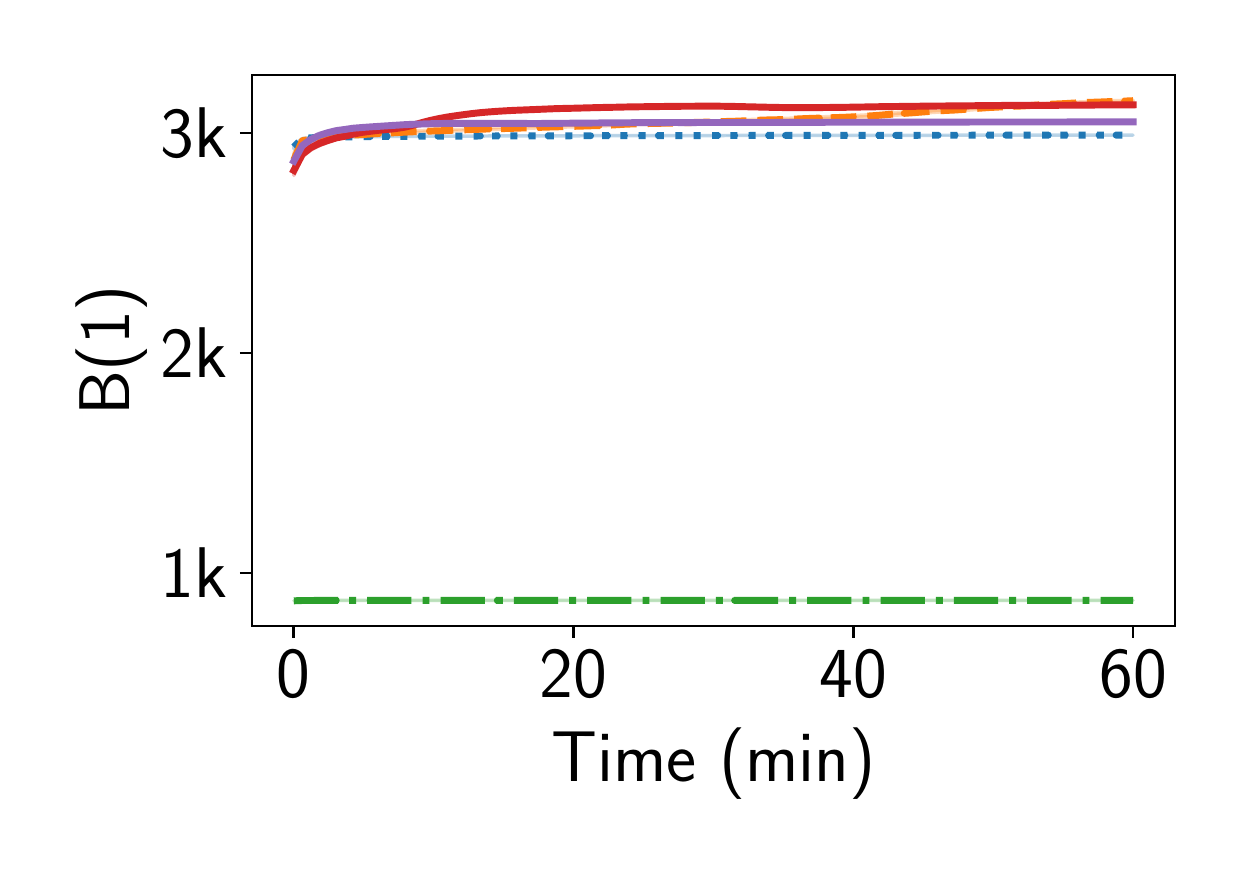}
		\includegraphics[width = \coverageFigureWidth]{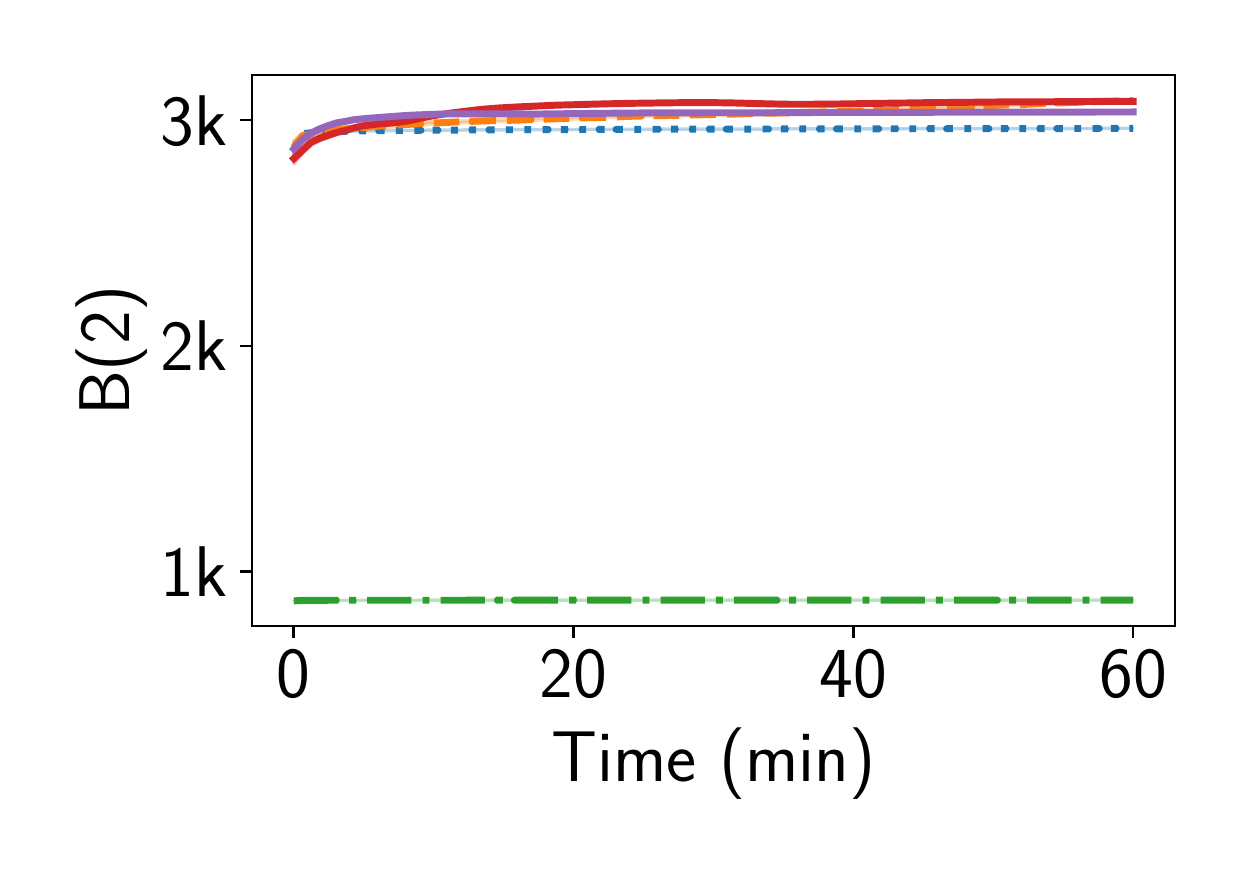}
	}\\[-2ex]
\end{tabular}
\caption{From left to right: Behavioral diversity of increasing order 0 to 2.}
\label{fig:evaluation_Hill_numbers}
\end{figure*}

In this section, we seek to answer \hyperref[sec:evaluation:RQ2]{RQ2}, that is,
whether \textsc{\ToolName} is able to generate test inputs that
have a higher behavioral diversity compared to the baseline techniques.
We measure the behavioral diversity by the \textsl{behavioral diversity index} $B(q)$ of order $q$,
defined in Equation \ref{eq:behavioral_diversity}.
In particular, we compare the behavioral diversity indices for $q \in \{0, 1, 2\}$, since
in the field of ecology, Hill numbers are usually reported for these orders as well.
Recall that the behavioral diversity $B(q)$ can be interpreted as follows:

\begin{itemize}
	\item[$q = 0$:] The total number of covered branches (i.e., branch coverage)
	\item[$q = 1$:] The effective number of typically executed branches
	\item[$q = 2$:] The effective number of commonly executed branches
\end{itemize}

\noindent The \textsl{effective number} of a set of covered branches can be seen as
the number of branches executed by the
proportionally same number of diverse inputs.
We are mostly interested in the results for $B(1)$ and $B(2)$,
since they emphasize the relative execution counts of
the typical and more common branches, respectively.

The results are shown in Figure \ref{fig:evaluation_Hill_numbers}.
From left to right, the columns depict the behavioral diversity of increasing order $q$ (from 0 to 2).
The first column shows for each subject and technique
the behavioral diversity of order 0, i.e.,
the total number of covered branches.
While RLCheck is able to generate the highest number of diverse valid inputs (Section~\ref{sec:results-RQ2}),
it performs the worst in terms of branch coverage.
When comparing the other approaches, 
both configurations of \textsc{\ToolName} have covered the highest number of branches
in three out of six benchmarks (Ant, Maven, Nashorn),
with \textsc{\ToolName-simple} additionally outperforming all other techniques in Tomcat.
However, Quickcheck outperforms \textsc{\ToolName} in the
two remaining benchmarks (Rhino and Closure).
Comparing against Zest, \textsc{\ToolName} is only significantly outperformed
in Closure, and also in Tomcat for \textsc{\ToolName-structure}.

The second column of Figure \ref{fig:evaluation_Hill_numbers} compares
the different techniques w.r.t. the $B(1)$ metric, i.e.,
the effective number of ''typically'' executed branches.
Here, a similar trend in the results can be observed.
While the relative performance of RLCheck remains the same,
both configurations of \textsc{\ToolName} perform
significantly better than all baseline approaches
in four out of six benchmarks (Maven, Rhino, Closure, Nashorn).
Most interestingly, when comparing against the plots of the first column 
($B(0)$, i.e., number of covered branches), \textsc{\ToolName} now
outperforms both Quickcheck and Zest in Rhino and Closure.
In the remaining benchmarks (Ant and Tomcat),
the performance of Zest is comparable to \textsc{\ToolName}.
Another observation is that
the $B(1)$ value of \textsc{\ToolName} tends to increase
over time, most noticeable on Ant, Rhino, and Closure.
This indicates that the inputs generated by \textsc{\ToolName}
not only diversely execute a fixed set of behaviors.
Instead, \textsc{\ToolName} is able to continuously increase the coverage of
new behaviors while maintaining the overall diversity of the triggered behavior.
In contrast, the $B(1)$ metric even slightly decreases over time
for Quickcheck in the Closure benchmark.
A possible explanation for this observation is that Quickcheck
is prone to disproportionately often trigger the most likely behaviors in the SUT.
On the other hand, most of the other covered branches may
have been executed only a few times by chance.
This increasing ''unevenness'' in executed behavior could consequently
result in a decline of the $B(1)$ value.

The third column shows the results for the $B(2)$ metric, i.e.,
the effective number of commonly executed branches.
Again, in four out of six benchmarks (Maven, Rhino, Closure, Nashorn), 
\textsc{\ToolName} performs significantly better than all baseline approaches.
For the Rhino and Closure benchmark,
the difference between \textsc{\ToolName} and the baselines
even increased compared to the results of the $B(1)$ metric (center column).
That is, both configurations of \textsc{\ToolName}
generate inputs that trigger ''common'' behavior
in a more diverse manner compared to the baselines.

Throughout the experiments, when comparing the two configurations of \textsc{\ToolName},
we have observed that \textsc{\ToolName-structure} generally performs better for shorter timeouts ($\leq 5$ minutes).
Since \textsc{\ToolName-structure} only saves inputs to the queue if they exhibit a new input structure,
the queue of saved inputs contains less, but more (structurally) diverse inputs.
As a result, \textsc{\ToolName-structure} initially explores more diverse behavior,
but the plain adaptive mutation strategy of \textsc{\ToolName-simple} seems to be sufficient to eventually
discover these behaviors as well.
Thus, the more explorative nature of \textsc{\ToolName-structure} may be more suitable for
contexts where runtimes are required to be short (e.g., property-based testing).

\subsection{RQ3: Finding Faults} \label{sec:results-RQ3}
Table \ref{fig:evaluation_bugs} shows the list of crashes that were discovered after a timeout of 24 hours,
deduplicated by benchmark and exception type (as done in the evaluations of Zest and RLCheck).
To answer \hyperref[sec:evaluation:RQ3]{RQ3}, we compare the different approaches concerning
the discovery times and the reliability of triggering a particular crash.
The results indicate that RLCheck performs the worst in terms of
fault finding capabilities, as this approach has triggered the least number of crashes.
In contrast, Zest performs the best as it has found one additional crash compared to
\textsc{\ToolName} and Quickcheck, though this crash was only found in $16\%$ of the trials.
This is is most likely due to the coverage-guided algorithm of Zest that allows to effectively explore
deeper paths in the semantic analysis stage compared to Quickcheck.
On the other hand, while \textsc{\ToolName} also leverages coverage guidance,
the approach focuses more on diversely executing many different behaviors through controlled mutations.
Thus, \textsc{\ToolName} may miss interesting edge cases that 
are more likely to be produced by completely random mutations.

Comparing the metrics for the bugs found by \textsc{\ToolName}, Zest, and Quickcheck, 
it can be noted that Quickcheck finds the crashes faster and more reliably.
The crashes in Closure and Rhino are typically found within minutes by Quickcheck,
while \textsc{\ToolName} and Zest require hours
to discover most of the crashes.
This result can potentially be explained by results in Section~ \ref{sec:results-RQ2}, which show that 
Quickcheck already achieved high coverage after a few minutes in these subjects.
In contrast, the search algorithms of \textsc{\ToolName} and Zest might have 
prioritized exploring other parts of the program, before eventually discovering the crash-inducing behaviors.
 
Interestingly, \textsc{\ToolName-structure} tends to find crashes faster than
\textsc{\ToolName-simple}, which suggests the importance
of the structural novelty heuristic for fault finding.

\begin{table*}[htpb]
\begin{tabular}{|l|c|c|c|c|c|} \hline
Crash-ID & BeDiv-simple & BeDiv-structure & Zest & Quickcheck & RLCheck \\ \hline
closure.StringIndexOutOfBoundsException & 645 (76\%) & 473 (73\%) & 82 (100\%) & 3 (100\%) & -\\
closure.NullPointerException & 425 (100\%) & 347 (100\%) & 149 (100\%) & 4 (100\%) & 7 (100\%)\\
closure.RuntimeException & 121 (100\%) & 88 (100\%) & 5 (100\%) & <1 (100\%) & 6 (100\%)\\
rhino.ArrayIndexOutOfBoundsException & - & - & 960 (16\%) & - & -\\
rhino.ClassCastException & 650 (40\%) & 371 (33\%) & 606 (56\%) & 192 (100\%) & -\\
rhino.IllegalStateException & 1 (100\%) & 2 (100\%) & 1 (100\%) & <1 (100\%) & <1 (100\%)\\
rhino.NullPointerException & 11 (100\%) & 7 (100\%) & 5 (100\%) & <1 (100\%) & -\\
rhino.VerifyError & 543 (80\%) & 452 (83\%) & 212 (100\%) & 6 (100\%) & 9 (100\%)\\
nashorn.AssertionError & 44 (100\%) & 48 (100\%) & 487 (66\%) & 125 (100\%) & -\\ \hline
\end{tabular}
\caption{Average time (in minutes) and reliability of triggering a particular crash.}
\label{fig:evaluation_bugs}
\end{table*}

\subsection{Threats to Validity}
\label{sec:eval-threats}
\noindent{\bf Internal Validity.} To avoid potential systematic errors that could pose threats to internal validity, we have designed our experiments (replication count, timeout, etc.) based on the guidelines provided by Klees et al. \cite{Klees2018EvaluateFuzzing}. Additionally, we have reused existing and available implementations of the baseline fuzzers with conforming parameter settings from the evaluation of RLCheck~\cite{reddy2020rlcheck}. The parameters for \textsc{\ToolName} are not tuned; thus we can provide a fair and realistic comparative evaluation of the different approaches.

\noindent{\bf External Validity.} Our evaluation focuses only on six programs (Ant, Maven, Rhino, Closure, Nashorn, and Tomcat) with two different input formats, namely XML and JavaScript. Whether our results can be generalized to other programs and other input formats is a threat to external validity. However, the programs under test represent complex, long-living, and mature programs with a widespread adoption. Thus, we argue that an application of \textsc{\ToolName} to similar programs would produce similar results.

\noindent{\bf Construct Validity.} The main question towards construct validity is whether the Hill numbers~\cite{Hill1973} for species diversity are actually a good metric to evaluate the behavioral diversity of the covered branches in a fuzzing campaign.
We argue that behavioral diversity is given not only if many different behaviors are covered (i.e., a high branch coverage), but each of the different behaviors is equally covered by many diverse inputs.
Utilizing Hill numbers as a metric specifically accounts for the possible variations in the diverse execution of different behaviors, which may differ greatly depending on the chosen fuzzing technique as shown in our evaluation.
\section{Related Work} \label{sec:rel-work}

{\bf Diversity in Fuzzing and Search-based Test Case Generation} There have been several fuzzing and search-based test case generation approaches that target diversity as one of their objectives. These approaches can be classified into approaches that a) aim to achieve diversity in the input space \cite{Biagiola0RT19,ChenLM04ART,soremekun2020inputs,VogelTG21,menendez2021hashing} or b) in the covered behavior of the program under test when executing the test cases. One of the earlier approaches in the first category is Adaptive Random Testing (ART)~\cite{ChenLM04ART}, a black box testing approach that aims to distribute the test cases over the entire input space. Technically, ART selects the next test case that maximizes the minimal distance to all already executed test cases. A similar approach is applied in the tool DIG~\cite{Biagiola0RT19} for testing web-based applications. The distance of test cases is computed based on the sequences of actions that traverse the navigational model of the web application. In search-based testing, Sapienz$^{div}$~\cite{VogelTG21} aims to maintain and improve a set of test cases for mobile applications that trigger diverse inputs, measured by the distance of test input sequences. The ``Uncommon inputs'' strategy from Soremekun et al.~\cite{soremekun2020inputs} creates inputs based on an inverted probabilistic grammar. If the original probabilistic grammar is learned from a set of common samples, the hypothesis is that inverting the probabilities would lead to uncommon and more diverse inputs.

\textsc{\ToolName} belongs to the second category of approaches since we focus on behavioral diversity and achieving input diversity as a byproduct. A first step towards behavioral diversity is taken in any greybox fuzzing approach (e.g., \cite{AschermannFHJST19NAUTILUS,afl, Bohme2016AFLFast, Bohme2017AFLGo,GanZQTLPC18CollAFL, Lemieux2018FairFuzz,Li2017Steelix, honggfuzz,libFuzzer}) that aim to maximize some coverage metric in the SUT, such as branch or statement coverage. The feedback loop in these greybox fuzzing approaches implements a novelty search that values test inputs that provide additional coverage. This feedback loop is also present in hybrid fuzzing approaches (e.g., \cite{StephensGrosenSallsEtAl2016,RawatJKCGB17VUzzer,Yun0XJK18QSYM}) that combine greybox fuzzing with symbolic execution~\cite{CadarDE08,Clarke1976,King1976,YangPRK14} or concolic testing~\cite{Sen2007}. Furthermore, the general idea of greybox fuzzing is taken a step further in FairFuzz~\cite{Lemieux2018FairFuzz}, VUzzer~\cite{RawatJKCGB17VUzzer}, and TortoiseFuzz~\cite{WangJLZBWS20TortoiseFuzz}.
FairFuzz is an AFL~\cite{afl} extension with the goal of triggering rare branches. The main idea of FairFuzz is to learn mutation masks and areas that have a higher chance of hitting these rare branches. TortoiseFuzz and VUzzer instead prioritize exploration of code regions with a high chance of containing a vulnerability. In the case of VUzzer, the goal is to cover error-handling code, and TortoiseFuzz aims to cover memory access operations.

In contrast to all these approaches, we argue that just hitting a branch once does not provide behavioral diversity.
The existing approaches basically aim to optimize the $B(0)$ metric (i.e., branch coverage) in Figure~\ref{fig:evaluation_Hill_numbers}, whereas \textsc{\ToolName} also provides behavioral diversity and scores well on the $B(1)$ and $B(2)$ metrics.

{\bf Fuzzing and Generation of Valid Inputs} The generation of syntactically and semantically valid inputs has always been the target of modern fuzzing approaches. Concerning syntactically correct inputs, the common approach is to use models to describe the input structure. Examples are input specifications \cite{Peach,JohanssonSLAG14T-Fuzz,KaksonenLT01} or grammars \cite{AschermannFHJST19NAUTILUS,eberlein2020egbf, Godefroid2008GrammarWhiteBoxFuzzing,Holler2012FuzzingCodeFragments, NguyenNKG20MoFuzz, soremekun2020inputs, Wang2019Superion}.
However, having just syntactically correct inputs is often not enough, and to explore deeper regions of the SUT, semantic validity of the inputs is required.
Zest~\cite{Padhye2019Zest} utilizes validity and code coverage feedback to produce inputs with high semantic coverage.
\textsc{\ToolName} is based on the same feedback mechanism as Zest, but extends it with novel structural mutation operators and a structure-aware fuzzing heuristic.
RLCheck~\cite{reddy2020rlcheck} uses reinforcement-learning to learn a policy to guide the generator towards high input diversity.
Similarly, \textsc{\ToolName} automatically adapts its mutation strategy based on the received feedback.
While our approach is built on top of these approaches, the main distinguishing factor is that we also aim to produce diverse behavioral inputs to have a more systematic exploration of the SUT's behavior.

\section{Conclusion and Future Work}
In this paper, we have described an approach to generate test inputs with high behavioral diversity.
That is, our approach does not only aim to \textsl{cover} as many behaviors as possible,
but also to \textsl{diversely execute} the different behaviors.
The key to our approach is to distinguish between \textsl{structure-changing mutations} that
allow to search for new behaviors triggered by specific input structures, and
\textsl{structure-preserving} mutations to diversely execute a particular behavior.
This method is complemented by an adaptive mutation strategy and a new
fuzzing heuristic that is based on the structural novelty of an input.
We implemented this approach in \textsc{\ToolName}, and show that it
outperforms the current state-of-the-art w.r.t. to a novel measure of behavioral diversity
that is inspired by a popular biodiversity metrics in ecology --- \textsl{Hill-numbers}.
In future work, we would like to provide guarantees for our approach by
evaluating measures such as the residual risk~\cite{BohmeLW21ResidualRisk}
and reliability of a SUT after we terminate a fuzzing campaign.

\begin{acks}
We would like to thank the anonymous reviewers for their insightful comments and suggestions, which helped to improve this paper. This research was partially funded by the Deutsche Forschungsgemeinschaft (DFG, German Research Foundation) – GR 3634/7-1 ProCI (421921612); GR 3634/6-1 FLASH (261444241); GR 3634/4-2 Emperor (392561203).
\end{acks}

\bibliographystyle{ACM-Reference-Format}
\bibliography{program_repair,diversity_ssbse,fuzzing,inputs-from-hell,symbolicexecution}
\end{document}

%% file: tikz_sets.tex
\tikzset{
  mycross/.pic={
    \draw[pic actions, line width=0.4mm] 
      (-5pt,0) -- (5pt,0)
      (0,-5pt) -- (0,5pt);
  },
}

\tikzset{
    inputspace/.pic={
    
    	\foreach \x in {0,1,2,...,10} {
	}
	
	\foreach \y in {0,1,2,...,10} {
	}
	
	\draw [<->,thick] (0,9.5) node (yaxis) [above] {$$}
        |- (10,0) node (xaxis) [right] {$$};
    
    	\node[font=\Huge\sffamily] at (0,5) [above, rotate=90] {Input Space};

	\node (v1) at (1,2) {};
	\node (v2) at (2,3.5) {};
	\node (v3) at (3.5,3) {};
	\node (v4) at (5,3.5) {};
	\node (v5) at (5.8,2.8) {};
	\node (v6) at (6.3,2) {};
	\node (v7) at (7.3,1.9) {};
	\node (v8) at (8.1,3.5) {};
	\node (v9) at (7.5,5.5) {};
	\node (v10) at (8.5,6.5) {};
	\node (v11) at (9.5,4) {};
	\node (v12) at (8.5,1) {};
	\node (v13) at (7,1.5) {};
	\node (v14) at (5.7,1) {};
	\node (v15) at (3.9,1.2) {};
	\node (v16) at (2.3,0.8) {};

	\begin{scope}[fill opacity=0.5]
	\filldraw[fill=black!30] ($(v1)$)
		to[out=90,in=180] ($(v2)$)
		to[out=0,in=180] ($(v3)$)
		to[out=0,in=180] ($(v4)$)
		to[out=0,in=90] ($(v5)$)
		to[out=270,in=180] ($(v6)$)
		to[out=0,in=270] ($(v8)$)
		to[out=90,in=270] ($(v9)$)
		to[out=90,in=180] ($(v10)$)
		to[out=0,in=90] ($(v11)$)
		to[out=270,in=0] ($(v12)$)
		to[out=180,in=0] ($(v13)$)
		to[out=180,in=0] ($(v14)$)
		to[out=180,in=0] ($(v15)$)
		to[out=180,in=0] ($(v16)$)
		to[out=180,in=270] ($(v1)$)
		;
	\end{scope}

	\foreach \v in {1,2,...,16} {
	}
	
	\node (w1) at (1,5.4) {};
	\node (w2) at (1.5,6.1) {};
	\node (w3) at (2.5,5.8) {};
	\node (w4) at (3.3,6.0) {};
	\node (w5) at (4.0,5.5) {};
	\node (w6) at (3.0,4.5) {};
	\node (w7) at (2.2,4.2) {};
	\node (w8) at (1.8,4.3) {};
	
	\begin{scope}[fill opacity=0.5]
	\filldraw[fill=black!30] ($(w1)+(0,0)$)
		to[out=90,in=180] ($(w2)+(0,0)$)
		to[out=0,in=180] ($(w3)+(0,0)$)
		to[out=0,in=180] ($(w4)+(0,0)$)
		to[out=0,in=90] ($(w5)+(0,0)$)
		to[out=270,in=0] ($(w6)+(0,0)$)
		to[out=180,in=0] ($(w7)+(0,0)$)
		to[out=180,in=270] ($(w1)+(0,0)$)
		;
	\end{scope}
	
	\foreach \w in {1,2,...,8} {
	}
	
	\node (x1) at (3.5,7.5) {};
	\node (x2) at (4.1,8.2) {};
	\node (x3) at (5.7,8) {};
	\node (x4) at (7.4,9) {};
	\node (x5) at (7.9,8.3) {};
	\node (x6) at (7.5,7.5) {};
	\node (x7) at (6.5,6.8) {};
	\node (x8) at (5.5,7.2) {};
	\node (x9) at (4.5,7) {};
	
	\begin{scope}[fill opacity=0.5]
	\filldraw[fill=black!30] ($(x1)$)
		to[out=90,in=180] ($(x2)$)
		to[out=0,in=180] ($(x3)$)
		to[out=0,in=180] ($(x4)$)
		to[out=0,in=90] ($(x5)$)
		to[out=270,in=90] ($(x6)$)
		to[out=270,in=0] ($(x7)$)
		to[out=180,in=0] ($(x8)$)
		to[out=180,in=0] ($(x9)$)
		to[out=180,in=270] ($(x1)$)
		;
	\end{scope}
	
	\foreach \x in {1,2,...,8} {
	}
	
	\node (y1) at (0.7,7.8) {};
	\node (y2) at (1.1,8.2) {};
	\node (y3) at (1.7,8.3) {};
	\node (y4) at (1.9,8.0) {};
	\node (y5) at (1.5,7.5) {};
	
	\begin{scope}[fill opacity=0.5]
	\filldraw[fill=black!30] ($(y1)$)
		to[out=90,in=180] ($(y2)$)
		to[out=0,in=180] ($(y3)$)
		to[out=0,in=90] ($(y4)$)
		to[out=270,in=0] ($(y5)$)
		to[out=180,in=270] ($(y1)$)
		;
	\end{scope}
}}

\tikzset{
    behaviorspace/.pic={
    
    	\foreach \x in {0,1,2,...,10} {
	}
	
	\foreach \y in {0,1,2,...,10} {
	}
	
	\draw [<->,thick] (0,10) node (yaxis) [above] {$$}
        |- (10,0) node (xaxis) [right] {$$};
    
    	\node[font=\Huge\sffamily] at (0,5) [above, rotate=90] {Behavior Space};
	\node[font=\LARGE\sffamily] at (0.1,9.5) [right] {Branch Hit Counts/Call Frequencies};
	\node[font=\LARGE\sffamily] at (0.1,0) [below right] {Likely Branches};
	\node[font=\LARGE\sffamily] at (9.9,0) [below left] {Rare Branches};
}}

%% file: macro.tex
\DeclareMathAlphabet{\mathpzc}{OT1}{pzc}{m}{it}

\makeatletter
\def\BState{\State\hskip-\ALG@thistlm}
\makeatother

\algnewcommand\algorithmicforeach{\textbf{for each}}
\algdef{S}[FOR]{ForEach}[1]{\algorithmicforeach\ #1\ \algorithmicdo}

\def\HiLi{\leavevmode\rlap{\hbox to \hsize{\color{gray!50}\leaders\hrule height .8\baselineskip depth .5ex\hfill}}} % highlighting lines in algorithms

\definecolor{light-gray}{gray}{0.8}
\definecolor{codegreen}{rgb}{0,0.6,0}
\definecolor{codegray}{rgb}{0.5,0.5,0.5}
\definecolor{codepurple}{rgb}{0.58,0,0.82}
\definecolor{backcolour}{rgb}{0.95,0.95,0.92}

\definecolor{javared}{rgb}{0.6,0,0} % for strings
\definecolor{javagreen}{rgb}{0.25,0.5,0.35} % comments
\definecolor{javapurple}{rgb}{0.5,0,0.35} % keywords
\definecolor{javadocblue}{rgb}{0.25,0.35,0.75} % javadoc
\definecolor{alg:gray}{gray}{0.85}

\theoremstyle{definition}
\newtheorem{definition}{Definition}[section]

\newcommand{\cmmnt}[1]{}

\let\othelstnumber=\thelstnumber
\def\createlinenumber#1#2{
    \edef\thelstnumber{%
        \unexpanded{%
            \ifnum#1=\value{lstnumber}\relax
              #2%
            \else}%
        \expandafter\unexpanded\expandafter{\thelstnumber\othelstnumber\fi}%
    }
    \ifx\othelstnumber=\relax\else
      \let\othelstnumber\relax
    \fi
}

\usepackage[colorinlistoftodos,prependcaption,textsize=tiny]{todonotes}

%% file: figure_1.tex
\begin{tikzpicture}[scale=0.45, transform shape]
	
	\pic{inputspace};
	
	\node[font=\Huge\sffamily] at (5,10) [] {(a) Traditional Fuzzer (AFL, libfuzzer,...)};
	
	\def\redpoints{
		(0.5,7), (1,6.9), (1.3,6.7), (1.5,6.7), (0.5,6.3), 
		(0.75,6.5), (1.3,3.6), (3,7.7), (3.3,8.1), (7,1.0), 
		(7.3,0.9), (8,6.7), (8.4,7.1), (8.6,7.4), (8.7,6.82), 
		(8.1,6.9), (7.8,7.2), (7.9,7.4), (8.0,7.3), (9.05,6.6), 
		(9.7,5.2), (7,6.1), (6,6.0), (4.9,5.8), (5.4,6.1), 
		(5.9,6.1), (5.6,5.7), (6.8,5.7), (5.3,5.5), (5.8,5.6), 
		(6.1,5.55), (6.5,5.35), (6.9,5.3), (5.45,5.1), (6.7,5.0), 
		(6.2,4.95), (5.8,4.8), (6.3,4.82), (7.3,4.73), (6.6,4.5), 
		(7.2,4.45), (7.4,4.), (7.05,4.65), (7.05,4.35), (6.8,4.2), 
		(6.9,4.02), (7.1,3.9), (6.5,4.1), (6.3,4.2), (6.1,3.9), 
		(6.2,3.78), (6.63,3.58), (6.98,3.3), (7,3.66), (7.68,3.47), (6.8,3), (7.15,2.86)
	}
	
	\foreach \p in \redpoints {
		\pic[line width=2pt,ACMRed,rotate=45] at \p {mycross}; 
  	}
	
	\def\greenpoints{
		(3.8,7.5), (3.8,5.5), (3.6,4.9), (3.1,2.6), (5.5,3), 
		(7.2,1.6), (9.2,5.54), (8.1,5.9), (8.5,6.1), (7.8,5.5), (8.3,5)
	}
	
	\foreach \p in \greenpoints {
		\pic[line width=2pt,black!30!green,rotate=45] at \p {mycross};
		%\node[fill=black,regular polygon, regular polygon sides=3,inner sep=1.5pt] at \p {};
  	}

\end{tikzpicture}
% Validity Fuzzer Graphic
\begin{tikzpicture}[scale=0.45, transform shape]

	\pic{inputspace};
	\node[font=\Huge\sffamily] at (5,10) [] {(b) Validity Fuzzer (Zest, RLCheck)};
	
	\pic[line width=2pt,ACMRed,rotate=45] at (7.4,6) {mycross};%
	\pic[line width=2pt,ACMRed,rotate=45] at (6.8,6.2) {mycross};%
	
	\def\greenpoints{
		(7,7.1), (6.4,7), (3.6,5), (5,3.35), (4,1.5),
		(9.3,2), (8.5,1.7), (9.11,5.71), (8.37,5.80), (8.62,4.99), 
		(9.10,5.20), (8.75,5.63), (8.73,5.79), (8.58,6.15), (8.81,5.32), 
		(8.55,5.63), (8.53,5.80), (8.67,5.09), (8.48,5.48), (8.70,5.76), 
		(8.42,5.80), (8.95,5.92), (8.45,5.55), (9.03,5.19), (8.73,5.51), 
		(8.41,5.01), (8.35,5.55), (8.92,5.13), (8.65,5.89), (9.29,5.60), 
		(8.64,5.74), (8.70,5.51), (8.57,5.95), (8.78,5.29), (8.64,5.83), 
		(7.99,5.36), (8.66,5.28), (7.92,5.54), (8.34,5.11), (8.46,4.96), 
		(8.24,5.59), (8.33,5.34), (8.54,5.17), (8.51,4.72), (8.19,5.61), 
		(8.15,5.22), (7.94,5.21), (8.13,6.01), (8.45,5.68), (8.33,5.47), 
		(8.39,5.62), (8.15,5.09), (8.04,5.17), (7.93,5.26), (8.20,5.18), 
		(8.26,5.24), (8.62,5.23), (8,5.8), (7.8,5.6), (7.7,5.83), (7.8,5.87),
		(8.62,4.84), (8.25,4.62), (8.23,4.83), (8.78,4.72), (8.41,4.86), 
		(8.31,4.82), (8.44,4.65), (8.68,4.79), (8.88,4.85), (8.89,4.66),
		(7.9,4.7), (9.2,4.84), (9.1,4.6)
	}
	
	\foreach \p in \greenpoints {
		\pic[line width=2pt,black!30!green,rotate=45] at \p {mycross};
		%\node[fill=black,regular polygon, regular polygon sides=3,inner sep=1.5pt] at \p {};
  	}
	
\end{tikzpicture}
% Behavioral diversity graphic
\begin{tikzpicture}[scale=0.45, transform shape]

	\pic{inputspace};
	\node[font=\Huge\sffamily] at (5,10) [] {(c) Behavioral Diversity (BeDivFuzz)};

	\pic[line width=2pt,ACMRed,rotate=45] at (7.4,6.3) {mycross};%
	\pic[line width=2pt,ACMRed,rotate=45] at (7.7,6.7) {mycross};%
	
	\def\greenpoints{
		(4.8,7.3), (4.8,7.3), (7.1,8), (7.3,8.3), (7.6,8.1), 
		(6.5,8.2), (6.7,7), (2,5.5), (2.8,5.7), (2,4.4), 
		(2.23,4.96), (3.51,5.05), (1.93,5.52), (2.42,5.00), 
		(2.70,5.27), (3.58,5.69), (3.14,5.04), (2.04,5.31), 
		(1.76,5.49), (2.03,4.98), (3.09,5.34), (2.78,4.72), 
		(2.13,5.66), (3.25,5.25), (3.4,4.80), (2,3), (2.4,2.80), 
		(3.4,4.80), (2.1,2), (3.5,2.3), (3.3,1.7), (3,2.2), (4,1.9), 
		(4.7,2.2), (4.8,1.6), (4.3,1.35), (5.2,1.9), (6,1.7),
		(8.2,1.8), (8.4,1.3), (9.1,1.5), (8.1, 6.0), (8.81,5.13), 
		(8.88,1.73), (8.70,1.62), (8.32,2.59), (8.35,2.10), 
		(8.92,5.61), (8.74,5.30), (8.72,5.48), (8.31,3.53), 
		(8.84,3.25), (8.98,1.62), (8.53,4.61), (9.08,5.45), 
		(8.48,1.68), (8.33,3.78), (8.80,1.55), (8.23,4.54), 
		(8.47,2.01), (8.71,2.84), (8.79,1.55), (8.56,5.30), 
		(8.22,4.86), (8.76,5.67), (8.97,5.13), (8.56,2.68), 
		(8.31,5.12), (8.67,5.20), (8.64,5.01), (8.20,5.76), 
		(8.26,4.81), (9.0, 4.0), (9.1, 2.7), (8.88, 2.1), (9.0, 4.0),
		(9.2, 4.4), (8.7, 4.25)
	}
	
	\foreach \p in \greenpoints {
		\pic[line width=2pt,black!30!green,rotate=45] at \p {mycross};
		%\node[fill=black,regular polygon, regular polygon sides=3,inner sep=1.5pt] at \p {};
  	}
	
\end{tikzpicture}

% Traditional Fuzzer Behavior 
\begin{tikzpicture}[scale=0.45, transform shape]
	\pic{behaviorspace};
	
	\def\heights{
		9, 9, 6.0, 5.6, 5, 3.2, 2.7, 1.3, 1.0, 1.0, 
		0.75, 0.5, 0.5, 0.5, 0.5, 0.5, 0.5, 0.25, 0.25, 0.25, 
		0.25, 0.25, 0.25, 0.25, 0.25, 0.25, 0.25, 0.25, 0.25, 0.25, 
		0.25, 0.25
	}
	
	\foreach \y [count=\x from 0] in \heights {
		\fill[opacity=0.3, fill=black] (0.25*\x, 0) rectangle ++(0.25, \y);
  	}	
\end{tikzpicture}
% Validity Fuzzer Behavior 
\begin{tikzpicture}[scale=0.45, transform shape]
	\pic{behaviorspace};
	
	\def\heights{
		8, 7, 6.7, 7.6, 4.7, 4, 3, 2.5, 1.5, 1.3, 
		0.5, 0.5, 0.5, 0.5, 0.5, 0.7, 0.7, 0.25, 0.25, 0.25,
		0.25, 0.5, 0.5, 0.25, 0.25, 0.25, 0.25, 0.25, 0.25, 0.25,
		0.25, 0.25, 0.25, 0.25, 0.25, 0.25, 0.25
	}
	
	\foreach \y [count=\x from 0] in \heights {
		\fill[opacity=0.3, fill=black] (0.25*\x, 0) rectangle ++(0.25, \y);
  	}	
\end{tikzpicture}
%BeDivFuzz
\begin{tikzpicture}[scale=0.45, transform shape]
	\pic{behaviorspace};
	
	\def\heights{
		7.2, 6.2, 6.5, 6.9, 2.5, 2.5, 1.5, 2.5, 2.5, 1.2,
		2.5, 2.5, 2.5, 2.5, 1.8, 2.5, 2.5, 2.5, 2, 2.5,
		2.8, 2.5, 2.5, 2.5, 3.3, 2.5, 2.5, 2.5, 2.5, 2.4,
		2.3, 1.3, 1, 0.7, 0.7, 0.5 
	}
	
	\foreach \y [count=\x from 0] in \heights {
		\fill[opacity=0.3, fill=black] (0.25*\x, 0) rectangle ++(0.25, \y);
  	}	
\end{tikzpicture}